\documentclass[pre,final,twocolumn,showkeys,showpacs,groupaddress,preprintnumbers,floatfix]{revtex4-1}

\usepackage{dynlearn}

\newcommand{\msym}{\meassymbol}
\newcommand{\MSym}{\MeasSymbol}

\newcommand{\ProcessLang}{L}
\newcommand{\domain}{\Lambda}
\newcommand{\radius} {R}
\newcommand{\neighborhood} {\eta}

\newcommand{\Localstate} {Local causal state\xspace}

\newcommand{\localstate} {local causal state\xspace}
\newcommand{\localstates} {local causal states\xspace}

\newcommand{\site} {r\xspace}
\newcommand{\point} {(\site, t)\xspace}

\newcommand{\lattice} {\mathcal{L}\xspace}
\newcommand{\stfield} [2] {\ensuremath{\mathbf{\msym}_{#1} ^{#2}}\xspace}
\newcommand{\STField} [2] {\ensuremath{\mathbf{\MSym}_{#1} ^{#2}}\xspace}

\newcommand{\stpoint} {\stfield{t}{\site}}
\newcommand{\STPoint} {\STField{t}{\site}}

\newcommand{\STPointprime} {\STField{t'}{\site'}}

\newcommand{\state}{\meassymbol}
\renewcommand{\State}{\MeasSymbol}

\newcommand{\plc}{\ell^-}
\newcommand{\PLC}{\mathtt{L}^-}
\newcommand{\flc}{\ell^+}
\newcommand{\FLC}{\mathtt{L}^+}

\newcommand{\causalfield} [2] {\ensuremath{\mathcal{S}_{#1} ^{#2}}\xspace}

\newcommand{\causalpoint} {\causalfield{t}{\site}}

\newcommand{\FullShift} {\MeasAlphabet ^ \mathbb{Z}}
\newcommand{\ShiftSpace} {\mathcal{X}}

\theoremstyle{definition}
\newtheorem{defn}{Definition}
\newtheorem{conj}{Conjecture}
\newtheorem{thrm}{Theorem}

\newtheorem*{prf}{Proof}
\newtheorem{lem}{Lemma}

\makeatletter
\newtheorem*{rep@thrm}{\rep@title}
\newcommand{\newreptheorem}[2]{
\newenvironment{rep#1}[1]{
 \def\rep@title{#2 \ref{##1}}
 \begin{rep@thrm}}
 {\end{rep@thrm}}}
\makeatother

\newreptheorem{thrm}{Theorem}

\newcommand{\lut}{\mathrm{LUT}\xspace}
\newcommand{\power}{n}
\newcommand{\excluded}{-}

\begin{document}

\def\ourTitle{
Spacetime Symmetries, Invariant Sets, and Additive Subdynamics\\
of\\
Cellular Automata
}

\def\ourAbstract{
Cellular automata are fully-discrete, spatially-extended dynamical systems that
evolve by simultaneously applying a local update function. Despite their
simplicity, the induced global dynamic produces a stunning array of
richly-structured, complex behaviors. These behaviors present a challenge to
traditional closed-form analytic methods. In certain cases, specifically when
the local update is additive, powerful techniques may be brought to bear,
including characteristic polynomials, the ergodic theorem with Fourier
analysis, and endomorphisms of compact Abelian groups. For general dynamics,
though, where such analytics generically do not apply, behavior-driven analysis
shows great promise in directly monitoring the emergence of structure and complexity in cellular automata. 
Here we detail a surprising connection between generalized symmetries in the spacetime fields of configuration orbits as revealed by the behavior-driven local causal states, invariant sets of spatial configurations, and additive subdynamics which allow for closed-form analytic methods. 
}

\def\ourKeywords{
coherent structures, spatially extended dynamical systems, emergence, symmetry breaking, cellular automata
}

\hypersetup{
  pdfauthor={James P. Crutchfield},
  pdftitle={\ourTitle},
  pdfsubject={\ourAbstract},
  pdfkeywords={\ourKeywords},
  pdfproducer={},
  pdfcreator={}
}

\title{\ourTitle}

\author{Adam Rupe}
\email{atrupe@ucdavis.edu}

\author{James P. Crutchfield}
\email{chaos@ucdavis.edu}

\affiliation{Complexity Sciences Center\\
Physics Department\\
University of California at Davis, One Shields Avenue, Davis, CA 95616}

\date{\today}
\bibliographystyle{unsrt}

\begin{abstract}
\ourAbstract
\end{abstract}



\preprint{\arxiv{1812.XXXXX}}

\title{\ourTitle}
\date{\today}
\maketitle

\setstretch{1.1}

\section{Introduction}
\label{sec:introduction}

Early systematic studies of cellular automata (CA) focused largely on
phenomenology \cite{Farm84a,Pack85b,Toff87a,Guto91b} and algebraic properties
of \emph{additive} cellular automata \cite{Mart84a,Lind84a}. Analysis was
possible due to the linear superposition exhibited by additive CAs. One
challenge was to extend the analytic techniques appropriate for linear CAs to
nonlinear CAs, which produce much richer and more physically-relevant
behaviors.

One approach restricted a nonlinear CA to a particular set of spatial
configurations that exercised only a linear subset of its behaviors. For
example, appropriately restricted the nonlinear CA rule 18 is equivalent to
linear CA rule 90 \cite{Gras83a}. That is, using one of these special
configurations as the initial condition, it does not matter if the evolution
follows rule 18 or rule 90, the resulting orbit is the same. This approach
was elaborated upon by Refs. \cite{Lind84a,Wolf83} and expanded upon by Ref.
\cite{Jen90b}, which proved nonlinear rules 126 and 146 can be reduced to the
linear rule 90. The latter also showed, under certain conditions, that more
general orbits of the nonlinear rules can be mapped to orbits of rule 90, then
mapped back; further extending the use of linear analytics on nonlinear rules.
Reference \cite{Moor97b} considered additional algebraic structures to
generalize additive CAs into a larger set of \emph{quasilinear} CAs that do not
obey superposition, but whose spacetime evolutions can still be predicted in
less time then general nonlinear CAs. More recently, Ref. \cite{Grav11a}
defined a special set of \emph{quasiadditive} spatial configurations for the
nonlinear rule 22 that emulate rule 146 at all even time steps. (Ref.
\cite{Jen90b} had already established that rule 146 emulates linear rule 90.)

Parallelling the earlier efforts, Crutchfield and Hanson developed rigorous
techniques that applied concepts from dynamical systems theory to cellular
automata behaviors \cite{Hans90a,Crut91d,Crut92a,Crut93a,Hans95a}: invariant
sets, stable and unstable manifolds, attractors, and basins. In particular,
they introduced techniques that automatically discovered a given CA's invariant
sets of spatial configurations \cite{McTa04a}. (Reference \cite{Nord88a} had
also identified invariant sets, but for a special CA class.) A CA's spacetime
shift-invariant sets of configurations are its \emph{domains}. They are key to
a CA's spatiotemporal organization: spacetime behaviors typically decompose
into domains and embedded defects---domain \emph{walls}, \emph{particles}, and
other \emph{coherent structures}.

More recently, we employed fully behavior-driven tools known as \localstates to
automatically discover (even hidden) spatiotemporal symmetries in CA behaviors
\cite{Rupe17b}. From this, coherent structures were shown to be localized
deviations from generalized symmetries of CA spacetime fields.

The following explores a surprising and subtle connection between these three
threads of analysis. It considers a CA at three levels of description: (i) its
global evolution function that updates spatial configurations, (ii) sets of
spatial configuration in the state space of all possible configurations, and
(iii) sets of spacetime-field orbits. Building on evidence presented in
Ref. \cite{Rupe17b}, we argue here that there is strong evidence that spacetime
field orbits resulting from evolution along an invariant set of spatial
configurations possess generalized spacetime symmetries that are captured by
the \localstates. This means domains can be equivalently described as invariant
sets in state space or as sets of behaviors---spacetime field orbits---with
generalized symmetries. Do domains similarly have a defining characteristic in
terms of the CA global evolution function? In fact, a wealth of examples do link
CA domains with additive subdynamics; nonadditive CAs can become additive when
evolving certain subsets of spatial configurations---domain invariant subsets.
However, this is not always the case. We present a counterexample: a CA that does
not become additive when evolving over a domain invariant set.

Defining a CA domain in terms of its equation of motion remains elusive.
Despite this, there is an interesting connection between CA domains, given as
either a spacetime invariant set or set of symmetric behaviors, and additive
subdynamics of the CA. After building up the necessary formalism, we revisit
examples given in the earlier literature showing nonlinear CAs displaying
linear behaviors to illustrate and explore when these linear behaviors are in
fact domain behaviors. The fact that many, \emph{but not all}, domains are
linear behaviors of CAs points to a more general theory still waiting to be
discovered. 

\section{Background}
\label{sec:background}

We first recall CAs and the subclass of additive CAs. We then review how
automata-theoretic methods are used to explore the temporal evolution of
configuration subsets. The notion of structured behavior that we use is then
laid out in our synopsis of spatiotemporal computational mechanics, including
how structures are detected by causal filtering and how to reconstruct a CA's
local causal states.

\subsection{Cellular automata}

A one-dimensional \emph{cellular automaton} or CA $(\MeasAlphabet^{\lattice},
\Phi)$ consists of a spatial lattice $\lattice  = \mathbb{Z}$ whose
\emph{sites} take values from a finite \emph{alphabet} $\MeasAlphabet$. A CA
\emph{state} $\state  \in \MeasAlphabet^{\mathbb{Z}}$ is the configuration of
all site values $\state^r \in \MeasAlphabet$ on the lattice. (For states
$\state$, subscripts denote time; superscripts sites.) CA states evolve in
discrete time steps according to the \emph{global evolution} $\Phi: \FullShift
\rightarrow \ShiftSpace \subseteq \FullShift$, where:
\begin{align*}
\state_{t+1} = \Phi(\state_t)
~.
\end{align*}
$\Phi$ is implemented through parallel, synchronous application of a \emph{local update rule} $\phi$ that evolves individual sites $\state_t^r$ based on their radius $\radius$ \emph{neighborhoods} $\neighborhood(\state^r) =$\\$ \{ \state^{r'} \; : \;  ||r-r'|| < \radius \}$:
\begin{align*}
\state_{t+1}^r = \phi \big(\; \neighborhood(\state_t^r)\; \big)
  ~.
\end{align*}

Stacking the states in a CA \emph{orbit} $\state_{0:t} = \{\state_0, \state_1,
\ldots, \state_{t-1}\}$ in time-order produces a \emph{spacetime field}
$\stfield{0:t}{} \in \MeasAlphabet^{\mathbb{Z} \otimes \mathbb{Z}}$.
Visualizing CA orbits as spacetime fields reveals the fascinating patterns and
localized structures that CAs produce and how the patterns and structures
evolve and interact over time.

\subsubsection{Elementary cellular automata}

The parameters $(\MeasAlphabet, R)$ define a CA \emph{class}. One simple but
nontrivial class is that of the so-called \emph{elementary} cellular automata
(ECAs) \cite{Wolf83} with a binary local alphabet $\MeasAlphabet = \{0,1\}$ and
radius $\radius = 1$ local interactions $\neighborhood(\state_t^r) =
\state_t^{r-1} \state_t^r \state_t ^{r+1}$. Due to their definitional
simplicity and wide study, we mostly explore ECAs in our examples.

A local update rule $\phi$ is generally specified through a \emph{lookup
table}, which enumerates all possible neighborhood configurations
$\neighborhood$ and their outputs $\phi( \neighborhood)$. The lookup table for
ECAs is given as:
\begin{align*}
\begin{tabular}{c c c | c}
\multicolumn{3}{c|}{$\eta$} & $O_\eta = \phi(\eta)$ \\
\hline
1 & 1 & 1 & $O_7$\\
1 & 1 & 0 & $O_6$ \\
1 & 0 & 1 & $O_5$ \\
1 & 0 & 0 & $O_4$ \\
0 & 1 & 1 & $O_3$ \\
0 & 1 & 0 & $O_2$ \\
0 & 0 & 1 & $O_1$ \\
0 & 0 & 0 & $O_0$
\end{tabular}
  ~,
\end{align*}
where each output $O_\eta = \phi(\eta) \in \MeasAlphabet$ and the $\eta$s are
listed in lexicographical order. There are $2^8 = 256$ possible ECA lookup
tables, as specified by the possible strings of output bits: $O_7 O_6 O_5 O_4
O_3 O_2 O_1 O_0$. A specific ECA lookup table is often referred to as an ECA
\emph{rule} with a \emph{rule number} given as the binary integer $o_7 o_6 o_5
o_4 o_3 o_2 o_1 o_0 \in [0,255]$. For example, ECA 172's lookup table has
output bit string $10101100$. 

In our analysis we examine the powers of the lookup table, also called
\emph{higher-order lookup tables}. The $\power^{\mathrm{th}}$-order lookup
table $\phi^\power$ maps the radius $\power \cdot \radius$ neighborhood of a
site to that site's value $\power$ time steps in the future. Said another way,
a spacetime point $\state_{t+\power}^r$ is completely determined by the radius
$ \power\cdot \radius$ neighborhood $\power$ time-steps in the past according
to:
\begin{align*}
\state_{t+\power}^r
  = \phi^\power\big (\neighborhood^{\power} (\state_t^r)\big)
  ~.
\end{align*}

\subsubsection{Additive CAs}

The special subclass of additive CAs is of particular interest. A CA is
\emph{additive} if the output $\phi(\neighborhood)$ may be written as a linear
combination of the neighborhood entries:
\begin{align}
\state_{t+1}^r = \phi \big(\; \neighborhood (\state_t^r)\; \big) = \sum_{i = -\radius}^{\radius} a_i \state_t^i \;  (\mathrm{mod} \; |\MeasAlphabet|)
~,
\label{eqn:additivity}
\end{align}
where $a_i \in \MeasAlphabet$.

Additive CAs are referred to as ``linear'' because they obey a linear
superposition principle. For any $N$ radius-$\radius$ neighborhood
configurations $\neighborhood_i$, the local update rule $\phi$ satisfies:
\begin{align}
\sum_{i=1}^N \phi(\neighborhood_i)
  = \phi \left(\sum_{i=1}^N \neighborhood_i\right)
  ~,
\label{eqn:linearity}
\end{align}
where $\sum$ denotes addition modulo $|\MeasAlphabet|$ over each site in the
neighborhoods \cite{Mart84a, Jen90b}. For the binary-alphabet CAs considered
here additivity and linearity are equivalent (each implies the other)
\cite{LeBr91a}. And so, we use both terms interchangeably.

Linear superposition enables powerful analytic techniques. For example,
representing spatial configurations via characteristic polynomials gives a
linear time-evolution via multiplication of a second polynomial representing
$\Phi$ \cite{Mart84a}. Or, one can employ the ergodic theorem for commuting
transformations together with Fourier analysis \cite{Lind84a}. Similarly,
ergodic theory interprets additive CAs as endomorphisms of a compact Abelian
group, which have been thoroughly investigated \cite{Lind77a}.

For general one-dimensional CAs with alphabet $\MeasAlphabet$ and neighborhood
radius $\radius$, there are $|\MeasAlphabet|^{|\MeasAlphabet|^{2\radius + 1}}$
update rules. However, only $|\MeasAlphabet|^{2\radius + 1}$ are linear.
Though there are cases where nonlinear rules may be mapped onto linear rules
\cite{Jen90b}, the analytic tools just described apply to a vanishingly small
subset of CA rules.

\subsection{Automata-theoretic CA evolution}

Rather than study how a CA evolves individual configurations, it is
particularly informative to investigate how CAs evolve sets of configurations
\cite{Hans90a,McTa04a}. This allows for discovery of structures in the state
space of a CA induced by $\Phi$.

For cellular automata in one spatial dimension, configurations $\state \in
\MeasAlphabet^{\mathbb{Z}}$ are strings over the alphabet $\MeasAlphabet$.
Sets of strings recognized by finite state machines are called \emph{regular
languages}. Any regular language $\ProcessLang$ has a unique minimal
finite-state machine $M(\ProcessLang)$ that recognizes or generates it
\cite{Hopc06a}. These automata are useful since they give a finite
representation of a typically infinite set of regular-language configurations.

Not all regular languages are appropriate as sets of spatial configurations for
cellular automata; we consider only sets that are subword closed and
prolongable in the sense that every word in the language can be extended to the
left and the right to obtain a longer word in the language. In automata theory
these are known as \emph{factorial, prolongable regular languages}. In symbolic
dynamics \cite{Lind95a} these are the languages of \emph{sofic
shifts}---closed, shift-invariant subsets of $\MeasAlphabet^{\mathbb{Z}}$ that
are described by finite-state machines. The language $\ProcessLang$ of a sofic
shift $\ShiftSpace \subseteq \MeasAlphabet^{\mathbb{Z}}$, a \emph{sofic
language}, is the collection of all words that occur in points $x \in
\ShiftSpace$. (See Definition 1.3.1 and Proposition 1.3.4 in Ref.
\cite{Lind95a} for details of sofic languages; sometimes called \emph{process
languages} in the computational mechanics literature.) Every state of the
machine $M(\ProcessLang)$ for a sofic language $\ProcessLang$ is both a start
and end state.

For the remainder of our development any language $\ProcessLang$ always refers
to a sofic language, and the machine $M(\ProcessLang)$ refers to the unique
minimal deterministic finite automaton of that language.

To explore how a CA evolves languages we establish a dynamic that evolves
machines. This is accomplished via finite-state transducers. Transducers are a
particular type of input-output machine that maps strings to strings
\cite{Broo89a}. This is exactly what a (one-dimensional) CA's global dynamic
$\Phi$ does \cite{Wolf84a}. As a mapping from a configuration $\state_t$ at
time $t$ to one $\state_{t+1}$ at time $t+1$, $\Phi$ is also a map on a
configuration set $\ProcessLang_t$ from one time to the next
$\ProcessLang_{t+1}$:
\begin{align}
\ProcessLang_{t+1} = \Phi(\ProcessLang_t)
	~.
\label{eq:CAMapsSets}
\end{align}

The global dynamic $\Phi$ can be represented as a finite-state transducer
$\mathrm{T}_\Phi$ that evolves a set of configurations represented by a
finite-state machine. This is the \emph{finite machine evolution} (FME)
operator \cite{Hans90a}. Its operation composes the CA transducer
$\mathrm{T}_\Phi$ and finite-state machine $M(\ProcessLang_t)$ to get the
machine $M_{t+1} = M(\ProcessLang_{t+1})$ describing the set
$\ProcessLang_{t+1}$ of spatial configurations at the next time step:
\begin{align}
M_{t+1} = \min \big( \mathrm{T}_\Phi \circ M(\ProcessLang_t) \big)
  ~.
\label{eq:FME}
\end{align}
Here, $\min (M)$ is the automata-theoretic procedure that minimizes the number
of states in machine $M$. While not entirely necessary for language evolution,
the minimization step is helpful when monitoring the complexity of
$\ProcessLang_t$. The net result is that Eq. (\ref{eq:FME}) is the
automata-theoretic version of Eq. (\ref{eq:CAMapsSets})'s set evolution
dynamic. Analyzing how the FME operator evolves configuration sets of different
kinds is a key tool in understanding CA emergent patterns.

\subsection{Computational Mechanics}

A question central to understanding CA spacetime behaviors concerns structure:
What spatial and spatiotemporal structures do CAs generate? How do they do so?
How are CA state spaces organized to support structures and their dynamical
emergence? Computational mechanics was developed to address the question of
structure in dynamical systems and stochastic processes \cite{Crut12a}.
We first review its pure-temporal version and then its extension to spacetime.

\subsubsection{Temporal processes and their causal states}

To identify and analyze pattern and structure in CA behaviors in a rigorous and
principled manner we employ the behavior-driven framework known as
\emph{computational mechanics} \cite{Crut88a,Crut12a}. Since it was most fully
developed in the temporal setting, consider first a stationary bi-infinite
stochastic process $\Process$---the distribution of all a system's allowed
behaviors or \emph{realizations} $\ldots \msym_{-2}, \msym_{-1}, \msym_0,
\msym_1, \ldots$ as specified by their joint probabilities $\Pr (\ldots,
\MSym_{-2}, \MSym_{-1}, \MSym_0, \MSym_1, \ldots)$. Here, $\MSym_t$ is the
random variable for the outcome of the measurement $\meassymbol_t \in
\MeasAlphabet$ at time $t$, taking values from a finite set $\MeasAlphabet$ of
all possible events. (Uppercase denotes a random variable; lowercase its
value.) We denote a contiguous chain of $\ell$ random variables as
$\MS{0}{\ell} = \MSym_0 \MSym_1 \dotsm \MSym_{\ell -1}$ and their realizations
as $\ms{0}{\ell} = \msym_0 \msym_1 \dotsm \msym_{\ell -1}$. (Left indices are
inclusive; right, exclusive.) We suppress indices that are infinite. A process
is \emph{stationary} when $\Pr \big(\MS{t}{t+\ell}\big) = \Pr
\big(\MS{0}{\ell}\big)$ for all $t$ and $\ell$.

The canonical object of computational mechanics, used to represent pattern and
structure in the behavior of a stochastic process, is the \emph{\eM}
\cite{Shal98a}. This is a type of stochastic finite-state machine known as a
hidden Markov model, which consists of a set $\CausalStateSet$ of \emph{causal
states} and transitions between them. The causal states are constructed for a
given process by calculating the equivalence classes determined by the
\emph{causal equivalence relation}:
\begin{align}
\ms{}{t} \; \CausalEquivalence \; \ms{}{t}^\prime \!\! \iff \!\!
  \Pr ( \MS{t}{} | \MS{}{t} \! = \! \ms{}{t} ) =
  \Pr ( \MS{t}{} | \MS{}{t} \! = \! \ms{}{t}^\prime )
  .
\label{eq:CausalEquiv}
\end{align}
In words, two pasts $\ms{}{t}$ and $\ms{}{t}^\prime$ are \emph{causally
equivalent}, i.e., belong to the same causal state, if and only if they make
the same prediction for the future $\Pr ( \MS{t}{} | \cdot)$. Equivalent states
lead to the same future conditional distribution. Behaviorally, the
interpretation is that whenever a process generates the same future (a
conditional distribution), it is effectively in the same state.

Each causal state $\causalstate \in \CausalStateSet$ is an element of the
coarsest partition of a process' pasts $\{\ms{}{t}: t \in \mathbb{Z}\}$ such
that every $\ms{}{t} \in \causalstate$ has the same predictive distribution:
$\Pr (\MS{t}{} | \ms{}{t}) = \Pr (\MS{0}{} | \cdot)$. The associated random
variable is $\CausalState$. The \emph{$\epsilon$-function} $\epsilon(\ms{}{t})$
maps a past to its causal state: $\epsilon: \ms{}{t} \mapsto \causalstate$. In
this way, it generates the partition defined by the causal equivalence relation
$\CausalEquivalence$. One can show that the causal states are the unique
\emph{minimal sufficient statistic} of the past when predicting the future
\cite{Shal98a}.

\subsubsection{Spatiotemporal processes, local causal states}

A spatiotemporal system, in contrast to a purely temporal one, generates a
process $\Pr(\ldots, \State_{-1}, \State_{0}, \State_{1}, \ldots)$ consisting
of the series of fields $\State_t$ over a spatial lattice $\lattice$.
A realization of a spatiotemporal process then is a \emph{spacetime
field} $\stfield{}{} \in \MeasAlphabet^{\lattice \otimes \mathbb{Z}}$,
consisting of a time series $\state_0, \state_1, \ldots$ of spatial
configurations $\state_t \in \MeasAlphabet^\lattice$.
$\MeasAlphabet^{\lattice \otimes \mathbb{Z}}$ is the orbit space of the process;
that is, time is added onto the system's state space. The associated spacetime
field random variable is $\STField{}{}$. A \emph{spacetime point} $\stpoint \in
\MeasAlphabet$ is the value of the spacetime field at coordinates
$\point$---that is, at location $\site \in \lattice$ at time $t$. The
associated random variable at that point is $\STField{t}{\site}$.

To extract pattern and structure from a system's spacetime field behaviors  we
employ a spatially local generalization of the causal equivalence relation Eq.
(\ref{eq:CausalEquiv}). For systems that evolve under a homogeneous local dynamic
and for which information propagates through the system at a finite speed, it
is quite natural to use lightcones as spatially local notions of pasts and
futures.

\begin{figure}[h]
\centering
\includegraphics{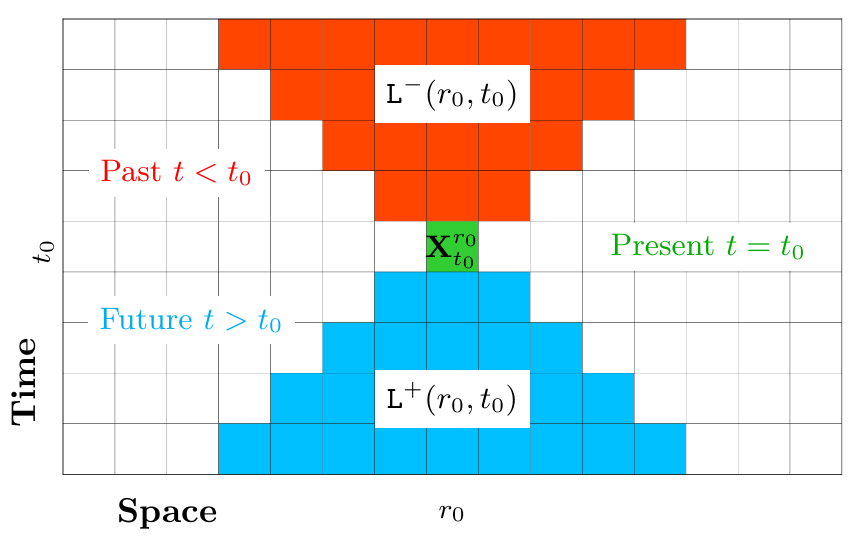}
\caption{Lightcone random variable templates: \textcolor{red}{Past lightcone
	$\PLC (r_0, t_0)$} and \textcolor{cyan}{future lightcone
	$\FLC (r_0, t_0)$} for \textcolor{green!67!black}{present spacetime
	point $\STField{t_0}{r_0}$} in a $1+1$ D field with nearest-neighbor (or
	radius-$1$) interactions.
	}
\label{fig:lightcones}
\end{figure}

Formally, the \emph{past lightcone} $\PLC$ of a spacetime random variable
$\STPoint$ is the set of all random variables at previous times that could
possibly influence it. That is:
\begin{align}
\PLC \point \equiv \big\{ \STPointprime :\; t' \leq t \; \mathrm{and} \;
||\site' - \site|| \leq c(t - t') \big \}
  ~,
\label{eq:plc}
\end{align}
where $c$ is the finite speed of information propagation in the system.
Similarly, the \emph{future lightcone} $\FLC$ is given as all the random
variables at subsequent times that could possibly be influenced by $\STPoint$:
\begin{align}
\FLC \point \equiv \big \{ \STPointprime :\; t' > t \; \mathrm{and} \;
||\site' - \site|| \leq c ( t' - t) \big \}
  ~.
\label{eq:flc}
\end{align}
We include the \emph{present} random variable $\STPoint$ in its past lightcone,
but not its future lightcone. An illustration for one-space and time ($1 + 1
$D) fields on a lattice with nearest-neighbor (or \emph{radius}-$1$)
interactions is shown in Fig~\ref{fig:lightcones}. We use $\PLC$ to denote the
random variable for past lightcones with realizations $\plc$; similarly, $\FLC$
those with realizations $\flc$ for future lightcones.

The choice of lightcone representations for both local pasts and futures is
ultimately a weak-causality argument: influence and information propagate
locally through a spacetime site from its past lightcone to its future
lightcone. As we will see in the rule 60 example in \cref{sec:exploring} A, the
utility of the \localstates as a behavior-driven tool is deeply rooted in this notion
of weak-causality.

Using lightcones as local pasts and futures, generalizing the causal
equivalence relation to spacetime is now straightforward. Two past lightcones
are causally equivalent if they have the same distribution over future
lightcones:
\begin{align}
\plc_i \; \CausalEquivalence \; \plc_j \iff \Pr \big(\FLC | \plc_i \big) = \Pr \big(\FLC | \plc_j \big)
\label{eq:LocalCEquiv}
  ~.
\end{align}
This \emph{local causal equivalence relation} over lightcones implements an
intuitive notion of \emph{optimal local prediction} \cite{Shal03a}. At some
point $\stpoint$ in spacetime, given knowledge of all past spacetime points
that could possibly affect $\stpoint$---i.e., its past lightcone
$\plc \point$---what might happen at all subsequent spacetime points that
could be affected by $\stpoint$---i.e., its future lightcone
$\flc \point$?

The equivalence relation induces a set $\CausalStateSet$ of \emph{local causal
states} $\causalstate$. A functional version of the equivalence relation is
helpful, as in the temporal setting, as it directly maps a given past
lightcone $\ell^-$ to the equivalence class $[\ell^-]$ of which it is a member:
\begin{align*}
\epsilon(\ell^-) & = [\ell^-] \\
  & = \{\ell^{-'}: \ell^- \sim_\epsilon \ell^{-'} \}
\end{align*}
or, even more directly, to the associated local causal state:
\begin{align*}
\epsilon(\ell^-) = \causalstate_{\ell^-}
  ~.
\end{align*}
The \localstates are the unique minimal sufficient statistics of past lightcones to predict future lightcones.

\subsubsection{Causal filtering}

As in temporal computational mechanics, the local causal equivalence
relation Eq. (\ref{eq:LocalCEquiv}) induces a partition over the space of
(infinite) past lightcones, with the \localstates being the equivalence
classes. We will use the same notation for local causal states as was used for
temporal causal states above, as there will be no overlap later:
$\CausalStateSet$ is the set of local causal states defined by the local causal
equivalence partition, $\CausalState$ denotes the random variable for a local
causal state, and $\causalstate$ for a specific realized causal state. The
$\epsilon$-function $\epsilon(\plc)$ maps past lightcones to their \localstates
$\epsilon :\plc \mapsto \causalstate$, based on their conditional distribution
over future lightcones.

For spatiotemporal systems, a first step to discover emergent patterns applies
the local $\epsilon$-function to an entire spacetime field to produce an
associated \emph{\localstate field} $\causalfield{}{} =
\epsilon(\stfield{}{})$. Each point in the \localstate field is a \localstate
$\causalpoint = \causalstate \in \CausalStateSet$.

The central strategy here is to extract a spatiotemporal process' pattern and
structure from the \localstate field. The transformation $\causalfield{}{} =
\epsilon(\stfield{}{})$ of a particular spacetime field realization
$\stfield{}{}$ is known as \emph{causal state filtering} and is implemented as
follows. For every spacetime coordinate $\point$:
\begin{enumerate}
\setlength{\topsep}{-3pt}
\setlength{\itemsep}{-3pt}
\setlength{\parsep}{-3pt}
\setlength{\labelwidth}{5pt}
\setlength{\itemindent}{0pt}
\item At $\stpoint$ determine its past lightcone $\PLC \point = \plc$;
\item Identify its local predictive distribution $\Pr(\FLC | \plc)$;
\item Determine the unique \localstate $\causalstate \in \CausalStateSet$
	to which it leads; and
\item Label the \localstate field at point $\point$ with $\causalstate$:
	$\causalpoint = \causalstate$.
\end{enumerate}
Notice the values assigned to $\causalfield{}{}$ in step 4 are simply the
labels for the corresponding \localstates. Thus, the \localstate field is a
\emph{semantic field}, as its values are not measures of any quantity, but
rather labels for equivalence classes of local dynamical behaviors as in the
\emph{measurement semantics} introduced in Ref. \cite{Crut91b}.

\subsubsection{Topological reconstruction}

Being a behavior-driven technique, the \localstates are reconstructed from
spacetime field realizations, rather than calculated from a system's equations
of motion. As the results presented in \cref{sec:exploring} below rely on
symmetries in \localstate fields, it is important to know whether one has a
faithful \localstate reconstruction or not.

Using the local causal equivalence relation as given in
Eq.~(\ref{eq:LocalCEquiv}) presents a challenge since the conditional
distributions over lightcones must be inferred from finite realizations. To
circumvent this, we instead use \emph{topological reconstruction}. This
replaces \emph{probabilistic morphs} $\mathrm{morph}_{\mathcal{P}}(\plc_i) =
\Pr \big(\FLC | \plc_i \big)$ with \emph{topological morphs}
$\mathrm{morph}_{\mathcal{T}}(\plc_i) = \{\mathrm{all} \; \flc_j \;
\mathrm{occurring \; with \;}\plc_i\}$, which are the supports of the
probabilistic morphs. Thus, two past lightcones are topologically (causally)
equivalent if they lead to the same \emph{set} of future lightcones. Contrast
this with being probabilistically (causally) equivalent, if they have the same
\emph{distribution} over future lightcones.

Topological reconstruction is particularly convenient for fully-discrete
systems such as CAs, since the topological morphs for finite-depth lightcones
can be \emph{exactly} reconstructed and the condition for topological
equivalence is \emph{exact}. (That is, are the topological morph sets the same
or not?) Moreover, the number of unique past lightcone-future lightcone
pairs seen in spacetime field data is monotone increasing, providing a
measure of convergence for identifying topological morphs. In short,
finite-depth approximations to the topological \localstates can be exactly
reconstructed with confidence.

For concreteness, \cref{sec:exploring}'s topological reconstruction of
\localstates uses past and future lightcone depths of $3$ for explicit symmetry
domains and past lightcone depth $8$ and future lightcone depth $3$ for hidden
symmetry domains. (Domain types are defined shortly.)

\section{Domains of Cellular Automata}
\label{sec:domains}

With CAs and the necessary analysis tools in hand, we now turn to explore their
``structured'' behaviors. In particular, we now consider behaviors called
\emph{domains}---terminology borrowed from condensed matter physics. Below we
give two distinct domain definitions; one in terms of dynamically-invariant
sets, the other in terms of local causal-state symmetries. For both, a formal
notion of domain is used to discover and describe these and other emergent
spacetime structures that form in CA spacetime fields.
Reference~\cite{Rupe17b} compared the structural analyses of CAs using both
domain definitions and reported on a strong empirical correspondence between
them. The results presented here in \cref{sec:exploring} further bolster this
correspondence, leading us to conjecture that the two definitions are
equivalent.

\subsection{Spacetime invariant sets}

The first systematic analysis of these and related structured behaviors of CAs
was done by Crutchfield, Hanson, and McTague
\cite{Hans90a,Crut91d,Crut92a,Crut93a,Hans95a,McTa04a,John10a}. Using the FME
operator, they discovered sets of spatially (and statistically) homogeneous
configurations that are invariant under a CA dynamic $\Phi$. 

Presently, we find it useful to restate and reinterpret these results using
symbolic dynamics \cite{Lind95a}. Recall that a \emph{shift space}, or simply
a \emph{shift}, $\ShiftSpace \subseteq \FullShift$ is a compact,
shift-invariant subset of the full-$\MeasAlphabet$ shift $\FullShift$. A \emph{point} $x = \ldots \msym_{-2},\msym_{-1}, \msym_0,
\msym_1, \ldots$ in a shift space is an indexed bi-infinite string of symbols
in $\MeasAlphabet$ and the \emph{shift operator} $\sigma$ increments the
indices of points by one; if $y = \sigma(x)$ for $x \in \ShiftSpace$, then $y_i
= x_{i+1}$ and by definition $y \in \ShiftSpace$. As the name suggests, a
\emph{sliding block code} $\Phi: \ShiftSpace \rightarrow \mathcal{Y}$ maps
points from one shift space to another using a sliding-window function $\phi$:
$y_i = \phi(x_{i-m:i+n})$, where $x \in \ShiftSpace$, $y \in \mathcal{Y}$. We
are particularly interested in surjective codes, also known as \emph{factor
maps}. The notational overlap with CA dynamics is intentional: CAs are uniform
sliding block codes ($m=n=R$) that commute with $\sigma$ \cite{Hedl69a}. 

We now give the spacetime invariant set definition of CA domains using this symbolic dynamics formalism. We consider sets of CA configurations given as shift spaces, and these are invariant sets of the CA if the CA dynamic $\Phi$ is a factor map from that shift space to itself. 
\begin{defn}
Consider a CA $\Phi$ and a set $\domain = \{\domain_1, \domain_2, \dots, \domain_{\widehat{p}}\}$ of shift spaces $\domain_i \subseteq \FullShift$. Together, this set of shifts is a \emph{domain} of $\Phi$ if the following hold:
\begin{enumerate}
\item \emph{Spatial invariance}: Each $\domain_i \in \domain$ is an
	\emph{irreducible sofic shift}. That is, the set of strings in each
	$\domain_i$ is generated by a strongly-connected finite-state machine
	$M(\domain_i)$.
\item \emph{Temporal invariance}: $\Phi: \domain_i \rightarrow
	\domain_{i+1 (\mathrm{mod} \; \widehat{p})}$ is a factor map from
	$\domain_i$ to $\domain_{i+1 (\mathrm{mod} \; \widehat{p})}$. Thus,
	$\Phi^{\widehat{p}}: \domain_i \rightarrow \domain_i$ is a factor from
	$\domain_i$ to itself, for all $\domain_i$. 
\end{enumerate}
\label{defn:dpiddomain}
\end{defn}
Each distinct $\domain_i$ is a \emph{temporal phase} of the domain and the
number $\widehat{p}$ of temporal phases is the \emph{recurrence time} of the
domain---the minimum number of time steps required for $\Phi$ to map a temporal
phase back to itself. The size $s$ of the minimal cycle in $M(\domain_i)$ is
the \emph{spatial period} of $\domain_i$. For all known examples, the spatial
period of each $\domain_i$ in a given $\domain$ is the same; thus, making $s$
the spatial period of the domain. 

An ambiguity arises here between $\Lambda$'s recurrence time $\widehat{p}$ and
its \emph{temporal period} $p$. For a certain class of CA domain (those with
explicit symmetries, see Sec.~\ref{sec:domains} C), the domain states $\state
\in \domain$ are periodic orbits of the CA, with orbit period equal to the
domain period: $\state = \Phi^p(\state)$. It is less clear how to define the
temporal period for domains in general using this formalism. The temporal
period appears to be related more to the \emph{spacetime shift spaces} of
domain orbits that results from evolving domain spatial shift spaces under
$\Phi$. The spacetime shift spaces of hidden symmetry domains are more
complicated objects than those of explicit symmetry domains. Notably, at
present, beyond particular cases it is not known how to generally relate the
domain spatial shifts to their resulting spacetime shifts. 

Given a CA $\Phi$, there are no general analytic solutions to
$\Phi^{\widehat{p}} (\domain_i) = \domain_i$. However, given a candidate shift
$\ShiftSpace$ it is computationally straightforward to find the factor
$\mathcal{Y}$ of $\ShiftSpace$ under $\Phi$ using the FME operator. That is, we
want to restrict the function-domain of $\Phi$ to $\ShiftSpace$ and then find
the set $\mathcal{Y}$ of images $y = \Phi(x)$ for all pre-images $x \in
\ShiftSpace$ so that $\Phi$ is a surjective map from $\ShiftSpace$ to
$\mathcal{Y}$. This is exactly what the FME operator does. If $\ShiftSpace$ is
an irreducible sofic shift and $\ShiftSpace = \Phi^{\widehat{p}}(\ShiftSpace)$
for some $\widehat{p}$, then $\ShiftSpace$ is a domain temporal phase of
$\Phi$. Since the FME evolves machines, we technically look for $M(\ShiftSpace)
= \Phi^{\widehat{p}}\bigl(M\left(\ShiftSpace\right)\bigr)$, where equality here
is given by machine isomorphism \footnote{Crutchfield and McTague implemented
an efficient, but exhaustive search algorithm to solve the invariant equation
using the enumerated library of machines of Ref. \cite{John10a}. Reference
\cite{Crut02a} analyzed ECA 22 using the approach.}. If $\widehat{p} > 1$, the
other temporal phases can be found using $\domain_{i + 1 (\mathrm{mod} \;
\widehat{p})} = \Phi(\domain_i)$.

\subsection{Local causal-state symmetries}

More recently, the local causal states were used to identify special CA
behaviors directly from spacetime fields \cite{Rupe17b}, rather than from
structures in state space that produce the spacetime fields, as with the
invariant sets just described. There, the causal filter $\causalfield{}{} =
\epsilon(\stfield{}{})$ was used to transform the time- and space-shift
invariance of a spacetime field $\stfield{}{}$ into explicit time and space
translation symmetries.

Consider a spatiotemporal process $\STField{}{}$, the set $\CausalStateSet$ of local causal states induced by the local causal equivalence relation $\sim_\epsilon$ over $\STField{}{}$, and the local causal state field $\causalfield{}{} = \epsilon(\stfield{}{})$ over the spacetime field realization $\stfield{}{}$. Let $\sigma_p$ denote the \emph{temporal shift operator} that shifts a spacetime field $\stfield{}{}$ by $p$ steps along the time dimension. This translates a point $\stpoint$ in the spacetime field as: $\sigma_p (\stfield{}{})_t^r = \stfield{t+p}{r}$. Similarly, let $\sigma^{s}$ denote the \emph{spatial shift operator} that shifts a spacetime field $\stfield{}{}$ by $s$ steps along the spatial dimension. This translates a spacetime point $\stpoint$ via: $\sigma^{s}(\stfield{}{})_t^r = \stfield{t}{r+s}$. 

\begin{defn}
A \emph{pure domain field} $\stfield{\domain}{}$ is a realization such that
$\sigma_{p}$ and $\sigma^{s}$ applied to $\causalfield{\domain}{} =
\epsilon(\stfield{\domain}{})$ form a symmetry group \cite{Holc82a}. The
generators of the symmetry group consist of the following translations:
\begin{enumerate}
\item
\emph{Temporal invariance}: For some finite time shift $p$ the domain causal state field is invariant:
\begin{align}
\sigma_p (\causalfield{\domain}{}) = \causalfield{\domain}{}
  ~,
\label{eq:LCSTempInvar}
\end{align}
and:
\item
\emph{Spatial invariance}: For some finite spatial shift $s$
the domain causal state field is invariant:
\begin{align}
\sigma^{s} (\causalfield{\domain}{}) = \causalfield{\domain}{}
  ~.
\label{eq:LCSSpatialInvar}
\end{align}
\end{enumerate}
The symmetry group is completed by including these translations' inverses,
compositions, and the identity null-shift $\sigma_\mathbf{0}(\stfield{}{})_t^r
= \stpoint$. The set $\CausalStateSet_\domain \subseteq\CausalStateSet$ is
$\domain$'s \emph{domain \localstates}: $\CausalStateSet_\domain = \{
\big(\causalfield{\domain}{}\big)_t^{\site} : ~ t \in \mathbb{Z}, \site \in
\lattice\}$.

A \emph{domain} $\domain$ of $\STField{}{}$ is the set of all realizations
$\stfield{\domain}{}$ such that $\causalfield{\domain}{} =
\epsilon(\stfield{\domain}{})$ contains only local causal states from
$\CausalStateSet_\domain$ and has the defining spacetime symmetries. The set
$\domain$ is a spacetime shift space---a closed, spacetime-shift invariant
subset of $\MeasAlphabet^{\mathbb{Z} \otimes \mathbb{Z}}$.
\label{defn:lcsdomain}
\end{defn}

The smallest integer $p$ for which the temporal invariance of
\cref{eq:LCSTempInvar} is satisfied is $\domain$'s \emph{temporal period}.  The
smallest $s$ for which \cref{eq:LCSSpatialInvar}'s spatial invariance holds is
$\domain$'s \emph{spatial period}.

The domain's \emph{recurrence time} $\widehat{p}$ is the smallest time shift
that brings $\causalfield{\domain}{}$ back to itself when also combined with
finite spatial shifts. That is, $\sigma^j
\sigma_{\widehat{p}}(\causalfield{\domain}{}) = \causalfield{\domain}{}$ for
some finite space shift $\sigma^j$. If $\widehat{p} > 1$, this implies there
are distinct tilings of the spatial lattice at intervening times between
recurrences. The distinct tilings then correspond to $\domain$'s \emph{temporal
phases}: $\domain = \{\domain_1, \domain_2, \dots, \domain_{\widehat{p}}\}$.
For systems with a single spatial dimension, like the CAs we consider here, the
spatial symmetry tilings are simply $(\causalfield{\domain}{})_t = \cdots w
\cdot w \cdot w \cdots = w^\infty$, where $w = (\causalfield{\domain}{})_t^{i:
i+s}$. Each domain phase $\domain_i$ corresponds to a unique tiling $w_i$.

We note that in contrast to the invariant-set approach of
\cref{defn:dpiddomain}, the temporal period of a domain is generally
well-defined using the local causal-state definition of domain in
\cref{defn:lcsdomain}. In fact, it is a defining property of
\cref{defn:lcsdomain}, whereas the recurrence time is the defining property of
\cref{defn:dpiddomain}. (Though, the recurrence time is still well defined using
the local causal-state approach.) This highlights a key distinction between
\cref{defn:dpiddomain} and \cref{defn:lcsdomain}; the invariant-set approach of
\cref{defn:dpiddomain} defines domains in terms of \emph{spatial shift spaces}
and their invariance under $\Phi$, whereas the local causal-state approach of
\cref{defn:lcsdomain} defines domains in terms of generalized symmetry
properties of \emph{spacetime shift spaces} that are spacetime field orbits of
$\Phi$. 

While \cref{defn:dpiddomain,defn:lcsdomain} are independent definitions of CA
domain that even differ in what mathematical objects are identified as domains,
we have found a strong empirical correspondence between these two definitions,
as shown in Ref.~\cite{Rupe17b} and here in \cref{sec:exploring}. We formalize
this correspondence as follows.

Consider a CA $\Phi$ and two domain sets $\domain^1$ and $\domain^2$.
$\domain^1$ is a set of spatial shifts that satisfy \cref{defn:dpiddomain} for
$\Phi$: each $\domain^1_i \in \domain^1$ is an irreducible sofic shift such
that $\Phi^{\widehat{p}}(\domain^1_i) = \domain^1_i$. Since $\domain^1$ is a
set of shift spaces that are themselves sets of spatial configurations
$\state_{\domain^1}$, we can simply think of $\domain^1$ as the set of all
configurations in the collection $\{\state_{\domain^1} \; :
\;\state_{\domain^1} \in \domain^1_i \in \domain^1\}$ of invariant spatial
shifts. $\domain^2$ is a spacetime shift space that satisfies
\cref{defn:lcsdomain} for $\Phi$: for each spacetime field orbit
$\stfield{\domain^2}{} \in \domain^2$ the local causal-state field
$\causalfield{\domain^2}{} = \epsilon (\stfield{\domain^2}{})$ is comprised of
states from $\CausalStateSet_{\domain^2}$ and is time- and space- translation
invariant: $\sigma_p(\causalfield{\domain^2}{}) = \causalfield{\domain^2}{}$
and $\sigma^s(\causalfield{\domain^2}{}) = \causalfield{\domain^2}{}$. We
conjecture the following bijective relationship between $\domain^1$ and
$\domain^2$.

\begin{conj}
For each configuration $\state_{\domain^1} \in \domain^1$, its orbit under
$\Phi$ is in $\domain^2$ and each spacetime field $\stfield{\domain^2}{} \in
\domain^2$ is the orbit, under $\Phi$, of a configuration in $\domain^1$.  That
is, first, for all $\state_{\domain^1} \in \domain^1$, there is
$\stfield{\domain^2}{} \in \domain^2$ such that $\stfield{\domain^2}{} =
\{\state_{\domain^1}, \Phi(\state_{\domain^1}), \Phi^2(\state_{\domain^1}),
\Phi^3(\state_{\domain^1}), \ldots\}$. And, second, for all
$\stfield{\domain^2}{} \in \domain^2$ there is $\state_{\domain^1} \in
\domain^1$ such that $\stfield{\domain^2}{} = \{\state_{\domain^1},
\Phi(\state_{\domain^1}), \Phi^2(\state_{\domain^1}),
\Phi^3(\state_{\domain^1}), \ldots\}$.
\label{conj:domainequiv}
\end{conj}

Taking this conjecture to be true, the following uses $\domain$ and ``domain''
to refer both to sets of invariant spatial shifts ($\domain^1$) and the set of
orbits of those spatial shifts ($\domain^2$).  

\subsection{CA domain classification: Explicit versus hidden symmetry}

CA domains fall into one of two classes: explicit symmetry or
hidden symmetry. In the \localstate formulation, a domain $\domain$ has
\emph{explicit symmetry} if the time and space shift operators $\sigma_p$ and
$\sigma^{s}$---that generate the domain symmetry group over
$\causalfield{\domain}{} = \epsilon(\stfield{\domain}{})$---also generate that
same symmetry group over $\stfield{\domain}{}$. That is, $\sigma_p
(\stfield{\domain}{}) = \stfield{\domain}{}$ and
$\sigma^{s}(\stfield{\domain}{} ) = \stfield{\domain}{}$. From this, we see the
following.

\begin{lem}
Every explicit symmetry domain configuration $\state_\domain \in \domain$ of a CA $\Phi$ generates a periodic orbit of that CA, with the orbit period
equal to the domain temporal period.
\label{lem:periodicorbits}
\end{lem}

\begin{prf}
This follows since time shifts of the spacetime field are essentially
equivalent to applying the CA dynamic $\Phi$: $\stfield{t+p}{} = \sigma_p
(\stfield{}{})_t$ and $\state_{t+p} = \Phi^p (\state_{t})$. Thus, if
$\state_\domain$ is any spatial configuration of a domain spacetime
field---$\state_\domain = (\stfield{\domain}{})_t$, for any $t$---then $\Phi^p
(\state_\domain) = \state_\domain$ if and only if $\sigma_p
(\stfield{\domain}{}) = \stfield{\domain}{}$.
\end{prf}

A \emph{hidden symmetry domain} is one for which the time and space shift
operators, which generate the domain symmetry group over
$\causalfield{\domain}{}$, do not generate a symmetry group over
$\stfield{\domain}{}$: $\sigma_p (\stfield{\domain}{}) \neq
\stfield{\domain}{}$ or $\sigma^{s}(\stfield{\domain}{} ) \neq
\stfield{\domain}{}$ or both.

Domain classification for the invariant-set formulation is similar. A domain
$\domain$ has explicit symmetry if its spatial configurations $\state_\domain$
are not just shift invariant but also translation invariant, so that $y =
\sigma^s(\state_\domain) = \state_\domain$, for all $\state_\domain \in \domain$. If $\domain$ has hidden symmetry it is still shift invariant: $y = \sigma^s(\state_\domain) \in \domain$, but it is not translation invariant: $y = \sigma^s(\state_\domain) \neq \state_\domain$. From the machine representation, $\domain$ is a hidden symmetry domain if $M(\domain)$ has any local branching in transitions between states. That is, if there is any state in $M(\domain)$ such that there is more than one transition leaving that state, then $\domain$ has hidden symmetry. Notably, hidden symmetry domains are associated with a level of stochasticity in their raw spacetime fields. We occasionally refer to these as \emph{stochastic domains}. 

The distinction between stochastic and explicit symmetry domains is key to
the additivity results to follow.

\section{Domains and additive subdynamics}
\label{sec:correspondence}

The first identification of what we now call a CA domain came from the
observation that the \emph{nonlinear} rule 18, when evolving certain
configurations, emulates the \emph{linear} rule 90 \cite{Gras83a}. As described
in more detail below, we know that the set of configurations over which rule 18
is equivalent to rule 90 is in fact the \emph{domain} of rule 18. Thus, in the
case of the nonlinear rule 18, its domain represents a subset of behavior that
is actually linear. Below we investigate whether there is a more general
connection between domains and linear behaviors of CAs, but first we formalize
the notion of a nonlinear CA having a subset of linear behaviors. 

\subsection{CA subdynamics}

Laying the groundwork for a CA's subdynamics requires clarity on what is
meant by the CA dynamic itself. The global dynamic $\Phi$ is implemented
through synchronous, parallel application of the local dynamic $\phi$. And so,
$\Phi$ and $\phi$ are closely related, but they should not be conflated. In
what follows, \emph{dynamic} refers to $\phi$ and, more specifically, to
its lookup table. However, since $\phi$ is the building block of $\Phi$,
restrictions on $\phi$ induce restrictions on $\Phi$. Specifically, these
constrain the configurations that we consider evolving under $\Phi$.

A subdynamic of $\phi$ then is determined by a subset of elements from its
lookup table. (This need not be a proper subset, we consider the full dynamic
$\phi$ to be a subdynamic.) To formalize this we must recast $\phi$'s lookup
table (LUT) as a set. We do this by considering elements of the set
$\lut(\phi)$ as tuples of neighborhood values and their outputs under $\phi$:
$\lut(\phi) = \{\big(\neighborhood, \phi(\neighborhood)\big):
~\text{for~all}~\neighborhood \in \MeasAlphabet^{2 \radius + 1}\}$. We also
need to consider higher-order lookup tables. The generalization to
$\lut(\phi^\power)$ is straightforward:
\begin{align*}
\lut(\phi^\power) = \{
  \big(\neighborhood^\power, \phi^\power(\neighborhood^\power)):
  ~\text{for~all}~ \neighborhood^\power \in \MeasAlphabet^{2\power \radius + 1}
  \}
~,
\end{align*}
where the $\neighborhood^\power$ runs through all length $2 \power \radius + 1$ words in $\MeasAlphabet$. 

We consider two methods for removing elements from $\lut(\phi^\power)$ to
create a subdynamic of $\phi^\power$. The first is related to additive
dynamics.

\subsubsection{Lookup table linearizations}

Recall that an additive local dynamic may be written in the form of
\cref{eqn:additivity}. For each $\state^i, \; i \in \{-R, -R+1, \ldots, R-1, R\}$,
in neighborhood $\eta$ there is a coefficient $a_i \in \MeasAlphabet$ such
that the output of $\phi$ for that neighborhood is given by
\cref{eqn:additivity}. Every linear rule then is specified by a length $2
\radius + 1$ vector $\mathbf{a}$ of these coefficients. For example, ECA rule
90 is given by $\mathbf{a}_{90} = (1,0,1)$, since $\state^i_{t+1} =
\phi_{90}(\state^{i-1}_t \state^i_t \state^{i+1}_t) = 1\state^{i-1}_t + 0\state^i +
1\state^{i+1}_t \; (\mathrm{mod} \; 2)$.

For a linear rule given as a coefficient vector, the vector $\mathbf{o} =
\mathbf{a} \mathbf{N}^{\mathrm{T}}$ gives the lookup table outputs, where
$\mathbf{N}$ is the matrix of neighborhood values, each row of $\mathbf{N}$ is
a neighborhood $\neighborhood$, and these are given in lexicographical order.
Thus, we can refer to linear rules via their coefficient vector $\mathbf{a}$, which 
will allow us to easily enumerate every additive CA in a given class. 

Consider an arbitrary lookup table $\lut(\phi_\alpha)$ and an additive lookup
table $\lut(\phi_\beta)$ in the same CA class.

\begin{defn}
Construct the \emph{$\phi_\beta$-linearization of $\phi_\alpha$}, denoted
$\phi_{\alpha \leftrightarrow \beta}$, by removing elements $(\neighborhood,
\phi_\alpha(\neighborhood))$ from $\lut(\phi_\alpha)$ if and only if
$\phi_\alpha(\neighborhood) \neq \phi_\beta(\neighborhood)$. That is, we
only keep elements in $\lut(\phi_\alpha)$ that are also in $\lut(\phi_\beta)$.
\end{defn}

Being linear, $\phi_\beta$ has a coefficient vector $\mathbf{a}_\beta$, which
is now also associated with $\phi_{\alpha \leftrightarrow \beta}$. For each element
$(\neighborhood, \phi_{\alpha \leftrightarrow \beta}(\neighborhood)) \in
\lut(\phi_{\alpha \leftrightarrow \beta})$ the output $\phi_{\alpha \leftrightarrow
\beta}(\neighborhood)$ is given by $\mathbf{a}_\beta \cdot \neighborhood$,
treating the neighborhood $\neighborhood$ as a vector and the dot product sum is performed mod $|\MeasAlphabet|$. $\beta$ may be given
as a CA rule number or as a coefficient vector; e.g., $\lut(\phi_{18
\leftrightarrow 90})$ is the same as $\lut(\phi_{18 \leftrightarrow (1,0,1)})$.

Generalizing to higher-order lookup tables, $\lut(\phi^\power_{\alpha
\leftrightarrow \beta})$ denotes keeping only elements in $\lut(\phi^\power_\alpha)$
if and only if $\phi^\power_\alpha(\neighborhood^\power) =
\phi^\power_\beta(\neighborhood^\power)$. At higher powers, there may be
linearizations that are not powers of a linear rule in the CA class. For
example, take $\phi_\alpha$ to be an ECA---CA class $(\MeasAlphabet
= \{0,1\}, \radius = 1)$. At the second power, we may want a linearization with
coefficient vector $\mathbf{a} = (0,1,1,1,1)$, which is not the second power of
a linear ECA. However, we still want this as a possible linearization of
$\lut(\phi^2_\alpha)$. In such cases, the coefficient vector will be used for
$\beta$: $\lut(\phi^2_{\alpha \leftrightarrow (0,1,1,1,1)})$.

Notice that the construction of $\lut(\phi_{\alpha \leftrightarrow \beta})$ is symmetric
between $\lut(\phi_\alpha)$ and $\lut(\phi_\beta)$ (via set intersection), and so $\lut(\phi_{\alpha \leftrightarrow \beta}) = \lut(\phi_{\beta \leftrightarrow \alpha})$. That being said, $\lut(\phi_{\alpha \leftrightarrow \beta})$ \emph{linearizes} $\phi_\alpha$ to $\phi_\beta$, while the opposite is not true; $\lut(\phi_{\beta \leftrightarrow \alpha})$ does not nonlinearize $\phi_\beta$. Any subset of an additive lookup table is necessarily also additive. 

\subsubsection{Language-restricted lookup tables}

Lookup table linearization, as described, is a procedure for defining a
particular CA subdynamic by focusing purely on the lookup table for that CA.
Such a subdynamic may not be realizable on a nontrivial set of spatial
configurations, though. For example, ECA rule 204 is the identity rule, which
is linear with coefficient vector $\mathbf{a}_{204} = (0,1,0)$. If we reduce
rule 18 to rule 204, the only neighborhoods in $\lut(\phi_{18 \leftrightarrow
204})$ are $000$ and $101$. The only configurations (longer than $3$) with only
these neighborhoods are all-$0$ configurations.

The motivation for the second subdynamic-construction method derives directly
from the \emph{configurational consistency} lacking in the $\phi_{18
\leftrightarrow 204}$ example. While lookup table linearization subdynamics are
constructed based by considering the outputs $\phi_\alpha(\neighborhood)$,
language-restricted subdynamics are constructed employing the neighborhoods
$\neighborhood$ themselves. 

Consider a sofic language $L$ (\cref{sec:background} B), its machine $M(L)$, and an
arbitrary CA rule $\phi_\alpha$.

\begin{defn}
Construct the \emph{$L$-restriction of $\phi_\alpha$}, denoted $\phi_{\alpha |
L}$, by removing elements $(\neighborhood, \phi_\alpha(\neighborhood))$
from $\lut(\phi_\alpha)$ if and only if $\neighborhood \notin L$.
\end{defn}

That is, consider each neighborhood $\neighborhood$ as a length $2\radius + 1$
word and only keep the neighborhoods that belong to the language $L$.
Operationally, if $\neighborhood \in L$, there exists a path in $M(L)$ such
that concatenation of output symbols along that path give the word
$\neighborhood$.

This easily generalizes to higher powers of $\lut(\phi_\alpha)$, as we can
consider the neighborhoods $\neighborhood^\power$ as words of length $2 \power
\radius + 1$, keeping elements $(\neighborhood^\power, \phi^\power_\alpha(\neighborhood^\power))$ in $\lut(\phi^\power_\alpha)$ if and only if $\neighborhood^\power \in L$.

\subsection{Domain-restricted lookup tables and their linearizations}

When a CA lookup table is restricted to one of its domain languages,
\emph{temporal consistency} is then added to the configurational consistency of
the language-restricted subdynamics. The surprising result is that a nonlinear
CA can have such a spatiotemporally-consistent subdynamic which is linear, thus
embedding realizable linear behavior in a generally nonlinear system. This is
true not only for the case of rule 18 and its domain, but we found that every
known ECA domain yields an additive subdynamic, as detailed below. This
naturally begs the question, are CA domains \emph{always} linear
behaviors---are domain-restricted lookup tables always additive? Further, as we
show below, linear CAs can produce \emph{only} domain behaviors---the
full-$\MeasAlphabet$ shift $\FullShift$ is invariant for any CA who's full
lookup table is additive. Another natural question then is whether or not
additive subdynamics always corresponds to evolution within an invariant set.
In short, long experience and much exploration leads one conjecture the general equivalence: 
\begin{align*}
\mathrm{Domain~invariant~set} \; \stackrel{?}{\iff} \; \mathrm{Additive~subdynamics}
  ~.
\end{align*} 

Perhaps not surprisingly, the connection between CA domains and additive
subdynamics turns out to be more subtle and complicated. Though all ECA domains
we know of correspond to additive subdynamics, we know of at least one CA in
the class $(\MeasAlphabet=\{0,1\}, R=2)$ with a nonlinear domain subdynamic.
This counterexample shows it cannot be generally true that CA evolution within
a domain invariant set must necessarily be linear evolution.

Given that there is no general equivalence, why is that so many domains
\emph{do} correspond to linear subdynamics? While we found a counterexample for
$\mathrm{CA} \; \mathrm{domain} \implies \mathrm{additive} \;
\mathrm{subdynamics}$, extensive exploration has yet to find a counterexample
for the converse. Thus, it remains an open question whether $\mathrm{additive}
\; \mathrm{subdynamics} \implies \mathrm{CA} \; \mathrm{domain}$. In what
follows we state these questions formally and explore them with many detailed
examples and several rigorous results.

\section{Exploring domains and additive subdynamics}
\label{sec:exploring}

Before beginning, a few comments about implementing the analysis tools are in
order. ECA lookup tables and their subdynamics are easily analyzed by hand, but
higher powers of the lookup tables quickly become unmanageable. Fortunately,
the subdynamic formalism just developed lends itself to automation. In the
following, this was implemented in Python, allowing for exact symbolic
exploration of higher-order lookup-table linearizations. For example, given the
set $\lut(\phi^n_{\alpha | L})$ we can enumerate all possible $2nR+1$
linearizations $\mathbf{a}$ to check whether $\lut(\phi^n_{\alpha | L})
\subseteq \lut(\phi^n_{\alpha \leftrightarrow \mathbf{a}})$. If so,
$\phi^n_\alpha$ linearizes to $\mathbf{a}$ over language $L$.

Similarly, topological reconstruction of the \localstates for each example was
also implemented in Python. Since the \localstate labels in causal filtering
are arbitrary, we use a simple, but arbitrary alphabetical labeling. To avoid
confusion, we emphasize that for different CAs there is no connection between
states labeled the same. For instance, we give causal filterings of ECAs 90 and
150 in Fig.~\ref{fig:90+150}, and in each of these there are
\localstates labeled $\mathsf{A}$. There is no relation, however, between state
$\mathsf{A}$ of rule 90 and state $\mathsf{A}$ of rule 150.

For simplicity we limit our discussions to one-dimensional CAs with binary alphabet $\MeasAlphabet = \{0,1\}$ and finite radius. Examples will mostly focus on elementary cellular automata $(\MeasAlphabet=\{0,1\}, R=1)$.

\subsection{Additive CAs produce only domains}

We begin our explorations with linear CAs and find that \emph{all} behaviors they produce are domains. From the invariant set perspective of domain, we state this formally.

\begin{thrm}
Every nonzero linear CA $\Phi_\beta$ is a factor map from the
full-$\MeasAlphabet$ shift to itself:
\begin{align*}
\Phi_\beta : \FullShift \rightarrow \FullShift
  ~.
\end{align*}
\label{thrm:surjectivity}
\end{thrm}

\paragraph*{Proof sketch} Using linear superposition, \cref{eqn:linearity}, and additivity, \cref{eqn:additivity}, a right inverse $\widetilde{\Phi}_\beta$ can be constructed for any linear $\Phi_\beta$ so that $\widetilde{\Phi}_\beta \bigl(\Phi_\beta \left(\state \right) \bigr) = \state$, for all $\state \in \FullShift$. See \cref{app:surjectivity} for the detailed proof. 

From \cref{thrm:surjectivity} we see that every orbit under $\Phi_\beta$ lies
in the domain invariant set since the invariant set is the set $\FullShift$ of
all spatial configurations. Similarly, this means $\Phi_\beta$ induces no
spatial restrictions on its images. Though we must note this result is for
bi-infinite spatial configurations $\state \in \FullShift$. If a linear CA
evolves configurations using other global topologies (most commonly a finite
ring topology) then it may not be surjective. For example, finite spatial
configurations with an odd number of sites with value $1$ are not reachable by
the linear ECA $\Phi_{90}$; there is no $y$ such that $x = \Phi_{90}(y)$ for
finite $x$ with an odd number of $1$s \cite{Mart84a}. Such unreachable states
are also known as \emph{Garden of Eden} states. 

What is the \localstate signature of additive CAs that only produce domain
behaviors? The answer is remarkably simple. When spacetime fields are filtered
with the local causal-equivalence relation, there is only a single \localstate.
That is, for \emph{every} spacetime field $\stfield{}{}$ produced by
$\Phi_\beta$, the associated \localstate field $\causalfield{}{} =
\epsilon(\stfield{}{})$ consists of a single state. Thus, the filtered field
has trivial time and space translation symmetries. Moreover, since there is
only one \localstate, this symmetry can never be broken. And so, all spacetime
fields of $\Phi_\beta$ are necessarily pure domain fields.

\begin{figure}[htp]
  \centering
  \includegraphics{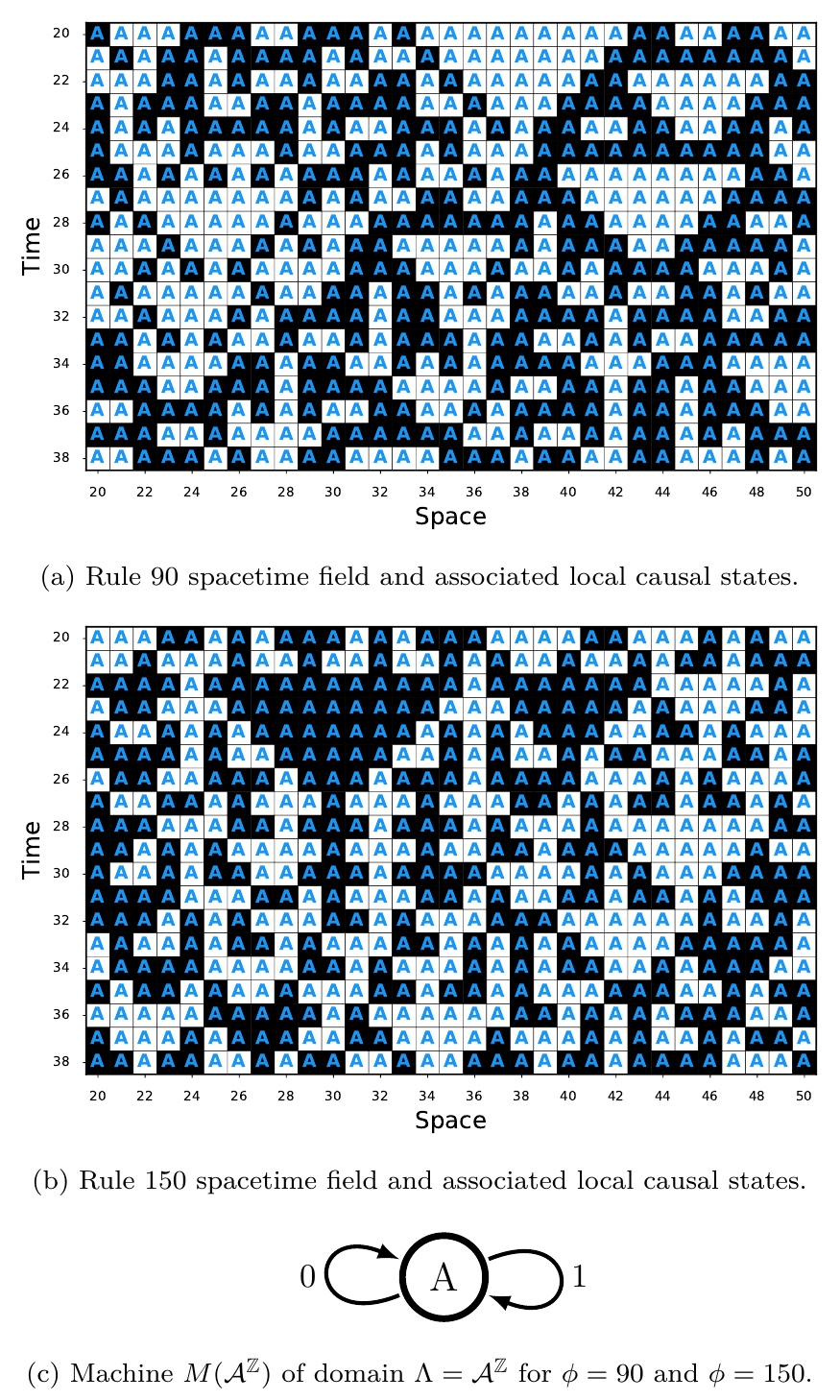}
\caption{Causally-filtered spacetime fields of (a) rule 90 and (b) rule 150,
	evolved from random initial conditions. White and black squares represent
	$0$ and $1$ CA site values, respectively. While blue letters are the
	associated \localstate label for that site. (c) Their domains
	$\domain_{90} = \FullShift$ and $\domain_{150} = \FullShift$ are described by the single-state
	machine $M(\FullShift)$.
	}
\label{fig:90+150}
\end{figure}

Figure~\ref{fig:90+150} displays spacetime fields $\stfield{}{}$ generated by
rule 90, $\mathbf{a}_{90}=(1,0,1)$, and rule 150, $\mathbf{a}_{150}=(1,1,1)$,
with the associated \localstate field $\causalfield{}{} =
\epsilon(\stfield{}{})$ superimposed on top. The white and black squares are
the site values $0$ and $1$, respectively, of the CA spacetime field. While the
blue letters denote the \localstates for each site in the field. Such diagrams,
with the \localstate field $\causalfield{}{} = \epsilon(\stfield{}{})$
superimposed over its associated CA spacetime field $\stfield{}{}$, are called
\emph{\localstate overlay diagrams}.

\subsubsection{Causal asymmetry of Rule 60}

\begin{figure}[htp]
 \includegraphics{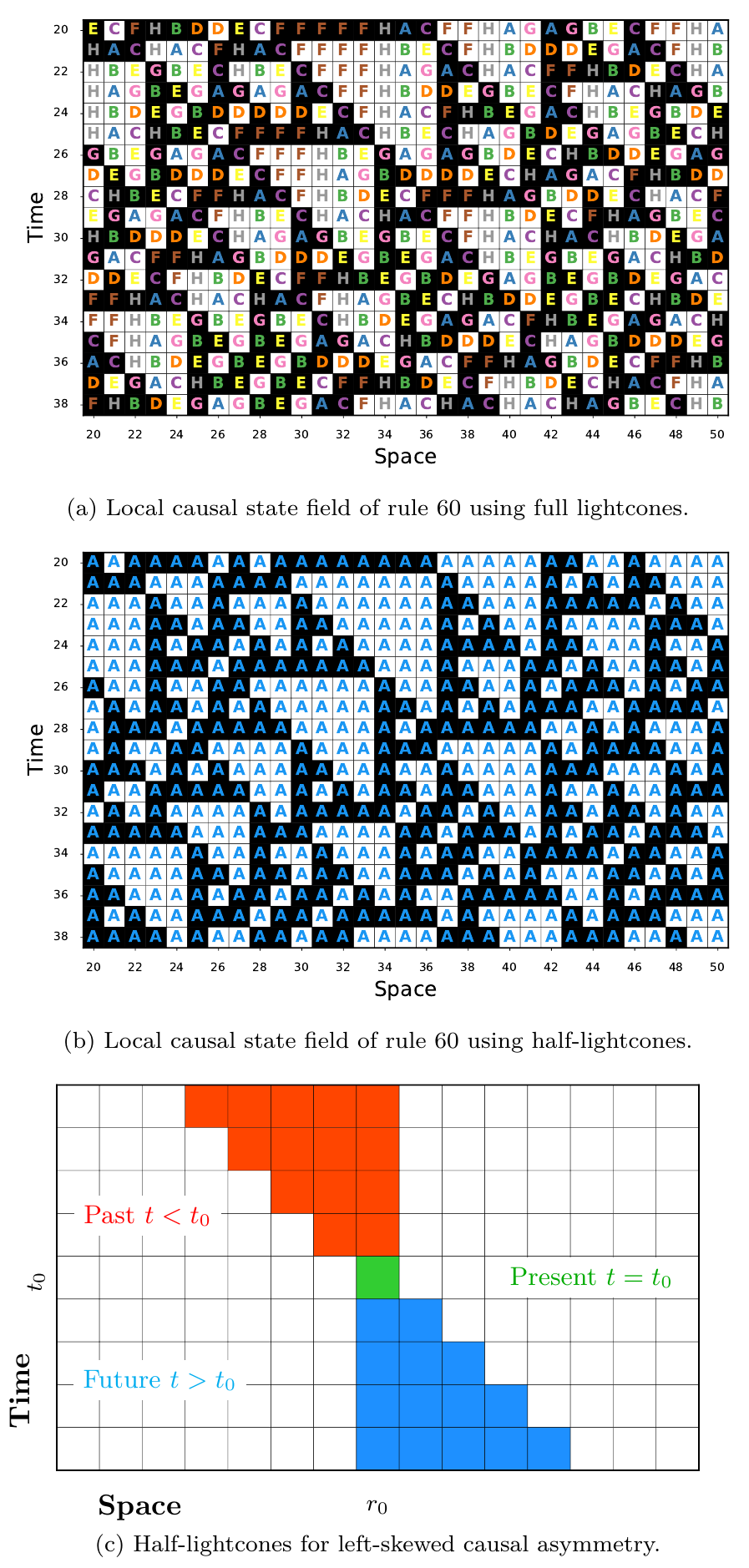}
\caption{Filtered spacetime field of rule 60, evolved from random initial
	conditions. White and black squares represent $0$ and $1$ site values,
	respectively. Colored letters denote the associated \localstate label for a
	site. (a) The result of using full lightcones, as depicted in
	Fig.~\ref{fig:lightcones}, for \localstate filtering. (b) \Localstate
	filtering appropriate to the left-skewed causal asymmetry of rule 60, using
	half-lightcones depicted in (c). This filtering recovers the single-state,
	trivially symmetric, \localstate field expected for additive CAs.
	}
\label{fig:rule60}
\end{figure}

While $\MeasAlphabet^{\mathbb{Z}}$ is invariant under rule 60, the \localstate
analysis reveals an interesting subtlety. Using standard lightcones, as
depicted in Fig.~\ref{fig:lightcones}, \localstate inference was run over
rule 60 using topological reconstruction. The resulting causal filtering is
shown in the overlay diagram of Fig.~\ref{fig:rule60}(a). As can be seen,
there are multiple causal states (cf. rules 90 and 150 above with one) and there
are no obvious spacetime translation symmetries in $\causalfield{}{}$.

This occurs since rule 60 breaks the causal symmetry assumption implicit in
standard lightcones. Since rule 60, $\mathbf{a}_{60}=(1,1,0)$, is the sum mod
$2$ of the center and left bit in the radius-$1$ neighborhood the right
neighbor is irrelevant to the dynamic. One takes this left-skewed causal
asymmetry into account by modifying the lightcone shape used in local causal
equivalence. The new, seemingly appropriate lightcone is depicted in
Fig.~\ref{fig:rule60}(c). Applying \localstate filtering using these
half-lightcones yields the overlay diagram of Fig.~\ref{fig:rule60}(b): A
single \localstate is revealed. Thus, taking into account the causal asymmetry,
\localstate analysis in fact demonstrates that rule 60 produces only
pure-domain fields. This holds similarly for rule 102,
$\mathbf{a}_{102}=(0,1,1)$, when the mirror-symmetry right-skewed
half-lightcones are used.

Since it demonstrates the consequences (and power) of the weak-causality
argument for using lightcones as local pasts and futures, the result here is
significant. Tracking weak-causality---how information locally propagates
through points in spacetime---is necessary for relating emergent behavior to
properties of the system that generated the behavior. Here, we know rule 60 is
additive and we know $\MeasAlphabet^{\mathbb{Z}}$ is invariant under
$\Phi_{60}$, so we know it should have a single \localstate. However, a single
\localstate is properly inferred only if we account for rule 60's inherent
causal asymmetry.

\subsubsection{Additive subdynamic generalization}

\Cref{thrm:surjectivity} shows that if $\lut(\phi_\beta)$ is additive, then
$\Phi_\beta$ is a factor from $\FullShift$ to itself. Consider though that
$\FullShift$ is in fact a sofic shift; its finite-state machine representation
is shown in \cref{fig:90+150} (c), and by definition $\lut(\phi_{\beta |
L(\FullShift)}) = \lut(\phi_\beta)$. So, an obvious generalization is to consider a general sofic shift $\ShiftSpace$, rather than $\FullShift$. 

Consider an arbitrary CA $\Phi_\alpha$ with $\lut(\phi_{\alpha |
L(\ShiftSpace)}) \subseteq \lut(\phi_{\alpha \leftrightarrow \beta})$; that is,
$\lut(\phi_{\alpha | L(\ShiftSpace)})$ is additive. Does this imply
$\Phi_\alpha$ is a factor map from $\ShiftSpace$ to itself: $\Phi_\alpha:
\ShiftSpace \rightarrow \ShiftSpace$? This cannot be the case, specifically
when $\Phi_\alpha$ is a linear CA. If $\phi_\alpha$ is additive, then
$\lut(\phi_{\alpha | L(\FullShift)})$ is additive. Thus, any subset of
$\lut(\phi_\alpha)$ will also be additive. In particular, $\lut(\phi_{\alpha |
L(\ShiftSpace)})$ will be additive for any sofic shift $\ShiftSpace$. This
would mean \emph{every} $\ShiftSpace$ would be an invariant set of
$\Phi_\alpha$. This is certainly not the case, however. For example,
$\ShiftSpace_{(0,\Sigma)}$ discussed below for rule 90 is not invariant for
rule 150.

This is not a concern, though, if we consider only nonlinear $\Phi_\alpha$.
It may then be the case that for a nonlinear $\Phi_\alpha$ and sofic shift
$\ShiftSpace$ that if $\lut(\phi^\power_{\alpha | L(\ShiftSpace)})$ is
additive, then $\Phi^\power_\alpha$ is a factor map from $\ShiftSpace$ to
itself: $\Phi^\power_\alpha : \ShiftSpace \rightarrow \ShiftSpace$. This though
is no longer strictly a generalization of \cref{thrm:surjectivity}. We note
that such a relation between nonlinear CAs $\Phi_\alpha$ and sofic shifts
$\ShiftSpace$ holds for all known cases, including those presented below.

\begin{conj}
Given an nonadditive $\Phi_\alpha$ and a sofic shift $\ShiftSpace$, if
$\lut(\phi^\power_{\alpha | L(\ShiftSpace)})$ is additive, then
$\Phi^\power_\alpha$ is a factor map from $\ShiftSpace$ to itself:
\begin{align*}
\Phi^\power_\alpha : \ShiftSpace \rightarrow \ShiftSpace
  ~.
\end{align*}
\end{conj}

The linear CAs, where the full dynamic is additive, produce only domain
behavior, as we saw. This leads to the more general question of when a
nonlinear CA being additive over a sofic shift $\ShiftSpace \subset \FullShift$
implies the CA is a factor map from $\ShiftSpace$ to itself. This remains an
open question, if somewhat more refined by the previous results.

Given this, we now turn to explore in a complementary direction. Does evolution
within a domain invariant set always correspond to linear behavior and an
additive subdynamic? If not, when does this happen? We start first with
explicit symmetry domains where we find that they do indeed always have a
corresponding additive subdynamic. In fact, they generally have many additive
subdynamics. Afterwards we examine the more complicated case of hidden symmetry
domains.

\subsection{Explicit symmetry domains}

Lemma~\ref{lem:periodicorbits} established that explicit symmetry domains are
periodic orbits of their CA. Due to this, explicit symmetry domains linearize
to the identity rule, which for ECAs is rule 204, $\mathbf{a}_{204} = (0,1,0)$.
For general $\MeasAlphabet = \{0,1\}$ CAs we denote the identity rule as
$\mathbf{a}_{\mathrm{I}}$, whose coefficient vector has a single $1$ for the
center bit and all other elements $0$.

Consider a CA $\Phi_{\alpha}$ with an explicit symmetry domain
$\domain_{\alpha}$ with temporal period $p$.

\begin{thrm}
The lookup table of $\phi_\alpha$, restricted to the domain $\domain_\alpha$,
linearizes to the identity rule at integer multiples of the domain temporal
period $p$; that is:
\begin{align*}
\lut(\phi^{\power p}_{\alpha | L(\domain_{\alpha})}) \subseteq \lut(\phi^{\power p}_{\alpha \leftrightarrow \mathbf{a}_{\mathrm{I}}})
~,
\end{align*}
for $\power = 1, 2, 3, \ldots$.
\end{thrm}

\begin{prf}
From Lemma~\ref{lem:periodicorbits}, any configuration
$\state_{\domain_{\alpha}}$ in the domain produces a periodic orbit of
$\phi_{\alpha}$, with orbit period $p$:
$\Phi^p_{\alpha}(\state_{\domain_{\alpha}}) = \state_{\domain_{\alpha}}$. Since
the full configuration returns after $p$ time steps, so do all the
individual sites of the configuration: $(\state_{\domain_\alpha})^{r_0}_{t_0 +
p} = (\state_{\domain_\alpha})^{r_0}_{t_0}$. Moreover,
$(\state_{\domain_\alpha})^{r_0}_{t_0 + \power p} =
(\state_{\domain_\alpha})^{r_0}_{t_0}$, for $\power = 1,2,3, \ldots$. Thus:
\begin{align*}
(\state_{\domain_\alpha})^{r_0}_{t_0 + \power p}
  & = \phi_\alpha^{\power p}\big(\neighborhood^{\power p}((\state_{\domain_\alpha})^{r_0}_{t_0})
\big) \\
  & = (\state_{\domain_\alpha})^{r_0}_{t_0} \\
  & = \mathbf{a}_{\mathrm{I}} \cdot
	\big(\neighborhood^{\power p}((\state_{\domain_\alpha})^{r_0}_{t_0}) \big)
  ~.
\end{align*}
This is an equivalent statement to $\lut(\phi^{\power p}_{\alpha | L(\domain_{\alpha})})
\subseteq \lut(\phi^{\power p}_{\alpha \leftrightarrow \mathbf{a}_{\mathrm{I}}})$.

\hfill $\blacksquare$
\end{prf}

For $\phi_{\alpha}$ with an explicit symmetry domain $\domain_{\alpha}$,
more linearizations are possible, based on the spacetime symmetries of
$\domain_{\alpha}$. If  $\domain_{\alpha}$'s recurrence time is smaller
than its temporal period $p$, it takes fewer time translations than $p$ to
return all sites to themselves if also paired with spatial translations. As the
\localstates fully capture the spatiotemporal symmetries of $\domain_{\alpha}$,
they are particularly convenient for expressing this.

Consider the \localstate field $\causalfield{}{} = \epsilon(\stfield{}{})$ of a
pure-domain field $\stfield{}{}$. From the definition of domain,
$\causalfield{t_0 + p}{r_0} = \causalfield{t_0}{r_0}$. And, for an explicit
symmetry domain this means $\stfield{t_0 + p}{r_0} = \stfield{t_0}{r_0}$, for
all $(r_0, t_0)$, which provides the connection to the identity rule
$\mathbf{a}_{\mathrm{I}}$ stated above. From the definition of recurrence time,
we have $\causalfield{t_0 + \widehat{p}}{r_0 + \delta} =
\causalfield{t_0}{r_0}$, for some spatial translation $\delta$. For explicit
symmetry domains this again implies the same invariance for $\stfield{}{}$.
There may be other spacetime translation symmetry generators such that
$\causalfield{t_0 + i}{r_0 + j} = \causalfield{t_0}{r_0}$. If
$\causalfield{t_0}{r_0}$ lies within the past lightcone of $\causalfield{t_0 +
i}{r_0 + j}$, there is a linearization of $\lut(\phi_{\alpha |
\domain_{\alpha}})$ at the $i^{\mathrm{th}}$ power, with linear coefficient
vector $\mathbf{a} = (a_{-i\radius}, a_{-i\radius+1}, \ldots, a_{-1}, a_0, a_1,
\ldots, a_{i\radius-1}, a_{i\radius})$ such that only $a_j = 1$ and all other
coefficients are zero.

\begin{figure}[htp]
\includegraphics{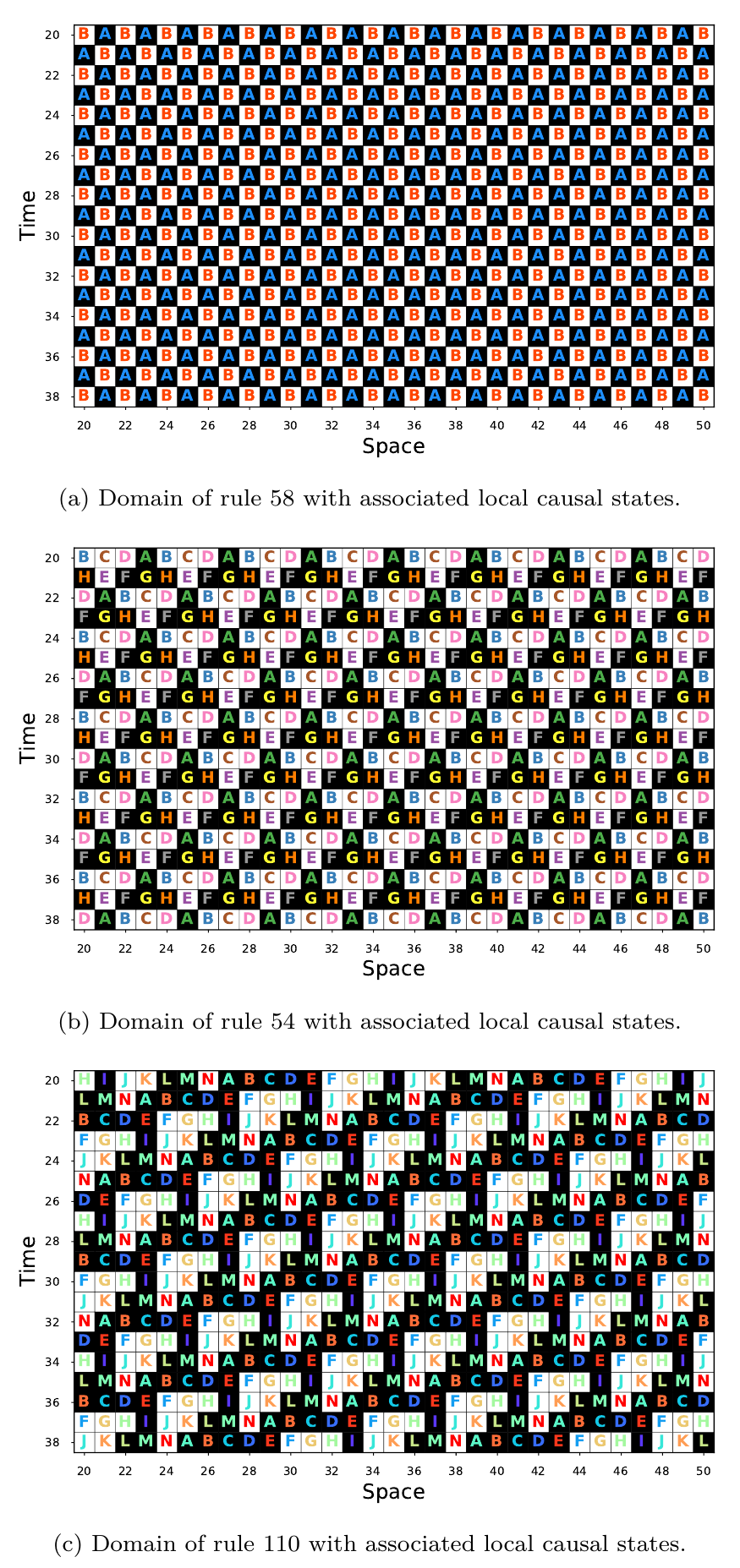}
\caption{Filtered spacetime fields from the explicit symmetry domains of rules
	58 (a), 54 (b), and 110 (c), with the associated \localstates superimposed
	on top.
	}
\label{fig:explicitsymmetries}
\end{figure}

For example, $\domain_{110}$ has recurrence time $\widehat{p} = 1$, temporal
period $p = 7$, and spatial period $s = 14$, as can be seen in
Fig.~\ref{fig:explicitsymmetries}(c). Thus, we know $\lut(\phi^7_{110 |
L(\domain_{110})}) \subseteq \lut(\phi^7_{110 \leftrightarrow 204})$. However,
we also see in Fig.~\ref{fig:explicitsymmetries}(c) that $\causalfield{t_0 -
3}{r_0 - 2} = \causalfield{t_0}{r_0}$. If we pick any site in the field, it
will have some \localstate label; e.g., {\sf A}. Shifting up three sites and
left two will always take one to same \localstate label. Exact symbolic
calculations showed that $\lut(\phi^3_{110 | L(\domain_{110})})$ linearizes
to $\mathbf{a} = (0,1,0,0,0,0,0)$. Similarly, $\causalfield{t_0 - 4}{r_0 + 2} =
\causalfield{t_0}{r_0}$ and $\lut(\phi^4_{110 | L(\domain_{110})})$ linearizes
to $\mathbf{a} = (0, 0, 0, 0, 0, 0, 1, 0, 0)$.

There may be yet further linearizations. These again derive from
$\domain_\alpha$'s symmetries, but the resulting linearizations are not related
to the identity rule. Let's start with an example. Consider again rule 110 and
its domain, as shown in Fig.~\ref{fig:explicitsymmetries}(c). We find several
linearizations of $\lut(\phi^7_{110 | L(\domain_{110})})$, including
$\mathbf{a}_{\mathrm{I}} = (0, 0, 0, 0, 0, 0, 0, 1, 0, 0, 0, 0, 0, 0, 0)$, the
identity rule discussed at the section's beginning. Another linearization we
find at the $7^{\mathrm{th}}$ power is $\mathbf{a} = (0, 1, 1, 1, 1, 1, 1, 0,
1, 1, 1, 1, 1, 1, 1)$. To understand this one, and others like it, it is again
useful to refer to spacetime diagrams of $\domain_{\alpha}$ ($\domain_{110}$ in
this case).

A linearization $\mathbf{a}$ means a spacetime point $\stfield{t}{r} \in \domain_\alpha$ is $0$ if $\neighborhood^\power(\stfield{t}{r}) \cdot \mathbf{a}$ is even and
$\stfield{t}{r}$ is $1$ if the dot product is odd, where we treat the
neighborhood as a vector. This must be satisfied at \emph{every} spacetime
point in a pure domain field $\stfield{\domain_\alpha}{}$. If:
\begin{align}
\stfield{t_0}{r_0} = \neighborhood^\power(\stfield{t_0}{r_0})
  \stackrel{(\mathrm{mod} \; 2)}{\cdot} \mathbf{a}
  ~,
\label{eq:ExplicitLinearization}
\end{align}
then for an explicit symmetry domain:
\begin{align*}
\stfield{t_0+ip}{r_0+js} =
\neighborhood^\power(\stfield{t_0+ip}{r_0+js}) \stackrel{(\mathrm{mod} \;
2)}{\cdot} \mathbf{a}
\end{align*}
for $i, j \in \mathbb{Z}$. Thus, there is only a relatively small number of
neighborhoods $\neighborhood^\power(\stfield{t}{r})$ in spacetime fields of
explicit symmetry domains that must satisfy \cref{eq:ExplicitLinearization}.
Said another way, the languages of explicit symmetry domains are very
restrictive. This means there are relatively few entries in
$\lut(\phi^n_{\alpha | L(\domain_\alpha)})$ that must obey additivity.

To further illustrate this linearization type, contrast the explicit symmetry
domains shown in Fig.~\ref{fig:explicitsymmetries}: (a) the rule 58 domain
$\domain_{58}$ with characteristics $\widehat{p} = 1$, $p = 2$, and $s = 2$;
and (b) the rule 54 domain $\domain_{54}$ with characteristics $\widehat{p} =
2$, $p = 4$, and $s = 4$. Now, compare their linearizations at the
$4^{\mathrm{th}}$ power, this being a multiple of the temporal period $p$ for
both domains.

The $4^{\mathrm{th}}$ power reveals more linearizations for $\domain_{58}$ than
for $\domain_{54}$ not related to the identity rule---those with more than one
nonzero element in the coefficient vector $\mathbf{a}$. Specifically, there are
$29$ linearizations for $\domain_{54}$ and $123$ for $\domain_{58}$. This is
expected, as $\domain_{58}$ is generated by smaller translations $s$ and $p$
than $\domain_{54}$. This means at a given power $\power$, there are fewer
distinct neighborhoods $\neighborhood^n$ in $\domain_{58}$ than in
$\domain_{54}$, and so there can be more linearizations $\mathbf{a}$ that can
satisfy \cref{eq:ExplicitLinearization} everywhere in the field.  

Before moving on to hidden symmetry domains, in closing we should highlight
the role of spacetime symmetries for linearizations of explicit symmetry
domains. Notice that the domain's spatial languages themselves were never
directly needed. Rather, the symmetries of the spacetime field orbits
and their orbit period (i.e., the domain temporal period) that were
key. This reflects the local causal-state perspective of domain, as in
\cref{defn:lcsdomain}, coming into play.

\subsection{Hidden symmetry domains}

In contrast, as we now show, linearizations of hidden symmetry domains are
based on the domain spatial languages and their recurrence times, as in
\cref{defn:dpiddomain}. Moreover, unlike explicit symmetry domains, we find
that not every hidden symmetry domain has a linearization. After going through
detailed examples of ECA domains, we will close with an $R=2$ CA that has a
``nonlinear'' domain.

\begin{figure}[htp]
 \centering
 \includegraphics{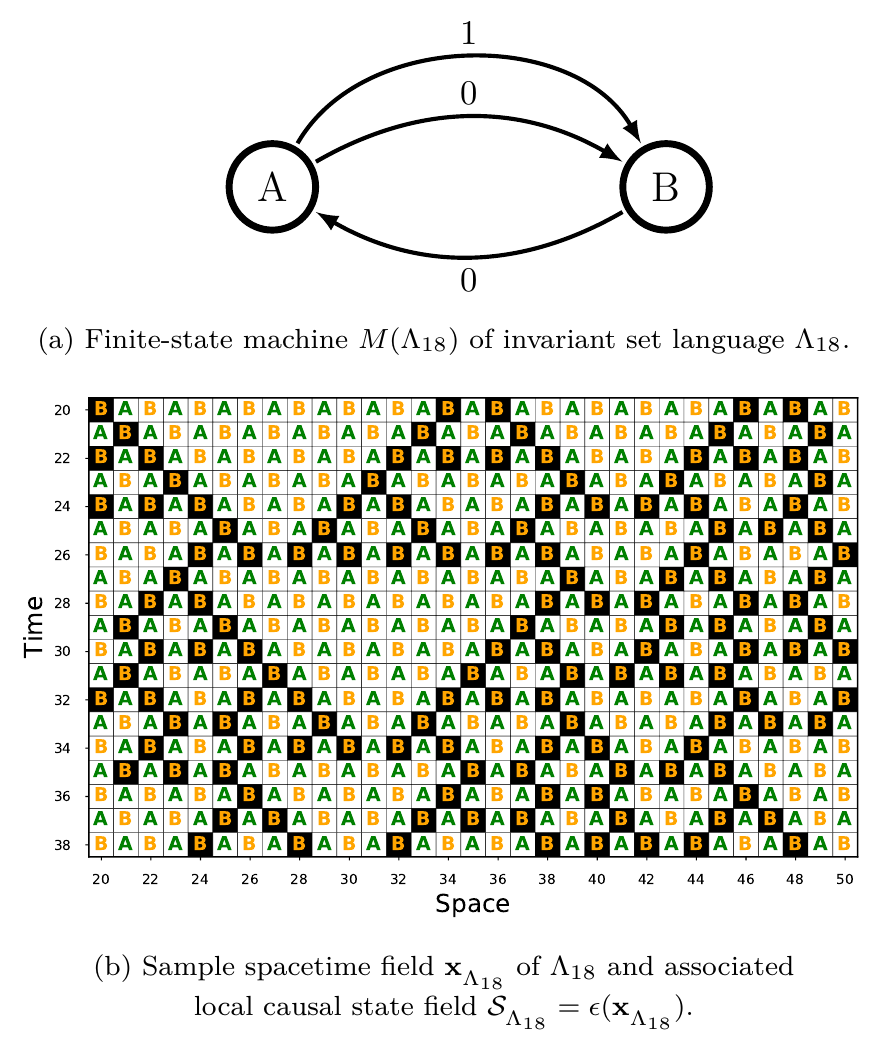}
\caption{(a) Finite-state machine $M(\domain_{18})$ for the invariant set
	language $L(\domain_{18})$ of the rule 18 domain. (b) Filtered spacetime
	field $\stfield{\domain_{18}}{}$ (white and black squares) of the rule 18
	domain $\domain_{18}$ with the associated \localstate field
	$\causalfield{\domain_{18}}{}=\epsilon(\stfield{\domain_{18}}{})$ (green
	and orange letters) superimposed.
	}
\label{fig:domain18}
\end{figure}

\subsubsection{Rule 18}

We start by examining in detail the original observations
\cite{Gras83a,Jen90b,Hans90a} concerning the domain in the nonadditive CA rule
18 and the domain's linearization to rule 90, $\mathbf{a}_{90} = (1,0,1)$. 
In words, $\phi_{90}$ updates lattice sites according to the sum mod $2$ of that
site's left and right neighbors:
\begin{align*}
\stfield{t+1}{r} & = \phi_{90}\big(\neighborhood(\stfield{t}{r})\big) \\
  & = \stfield{t}{r-1} + \stfield{t}{r+1} \; (\mathrm{mod} \; 2)
  ~.
\end{align*}
Table~\ref{tab:rule18} gives rule 90's lookup table.

Rule 18 is not additive, as can be seen from its lookup table also given in
Table~\ref{tab:rule18}. However, there are special behaviors produced by rule
18 that were originally noted due to the equivalence between these behaviors of
rule 18 and rule 90. More importantly, they suggested that a \emph{nonlinear
rule is capable of producing linear behaviors}. From Refs.
\cite{Hans90a,Crut91d,Crut92a,Crut93a,Hans95a}, we know the special behaviors
of rule 18 that emulate rule 90 are, in fact, rule 18's domain behaviors.

Rule 18's domain is the set of spatial configurations that is invariant under
$\Phi_{18}$ and their spacetime field orbits. This invariant set is the single
sofic shift $\domain_{18} = \{\ShiftSpace_{(0,\Sigma)}\}$, where $\Sigma$
represents wildcard-sites that can be either $0$ or $1$. Its domain language is
$L(\domain_{18}) = (0 \Sigma)^* + (\Sigma0 )^*$. The set's finite-state machine
$M(\domain_{18})$ is shown in Fig.~\ref{fig:domain18}(a). Since the machine
states lie in a single recurrent component, i.e., it has a single temporal
phase, the recurrence time of $\domain_{18}$ is $\widehat{p} = 1$. Its spatial
period is $s = 2$, since this is the size of the minimal cycle of
$M(\domain_{18})$.

Evolving spatial configurations $\state \in \domain_{18}$ creates spacetime
fields---their orbits $\stfield{\domain_{18}}{}$. Applying the causal
equivalence relation over these fields yields two \localstates, corresponding
to the fixed-$0$ and wildcard sites. A sample spacetime field
$\stfield{\domain_{18}}{}$ of $\domain_{18}$ and its causal filtering
$\causalfield{\domain_{18}}{} = \epsilon(\stfield{\domain_{18}}{})$ are shown
as a \localstate overlay diagram in Fig.~\ref{fig:domain18} (b). State
$\mathsf{A}$ corresponds to the fixed-$0$ sites and state $\mathsf{B}$ the
wildcard states. These states appear in a checkerboard tiling in the field,
displaying the defining spacetime symmetry of $\domain_{18}$. At each time step
the same two states tile the spatial lattice, giving the recurrence time
$\widehat{p} = 1$. The spatial period $s = 2$ and temporal period $p = 2$ are
found from $\causalfield{\domain_{18}}{}$'s space and time translation
invariance.

\begin{table}
\begin{tabular}{c c c | c | c | c | c}
\multicolumn{3}{c|}{$\eta$} & $\phi_{90}(\eta)$ & $\phi_{18}(\eta)$ & $\phi_{18 \leftrightarrow 90}(\eta)$ &$\phi_{18 | L(\domain_{18})}(\eta)$ \\
\hline
1 & 1 & 1 & 0 & 0 & 0 & $\excluded$\\
1 & 1 & 0 & 1 & 0 & $\excluded$ & $\excluded$\\
1 & 0 & 1 & 0 & 0 & 0 & 0 \\
1 & 0 & 0 & 1 & 1 & 1 & 1\\
0 & 1 & 1 & 1 & 0 & $\excluded$ & $\excluded$\\
0 & 1 & 0 & 0 & 0 & 0 & 0\\
0 & 0 & 1 & 1 & 1 & 1 & 1\\
0 & 0 & 0 & 0 & 0 & 0 & 0
\end{tabular}
\caption{Lookup tables for rule 90 ($\phi_{90}$) and rule 18 ($\phi_{18}$) as
	well as for rule 18 linearized to rule 90 ($\phi_{18 \leftrightarrow 90}$)
	and rule 18 restricted to its domain ($\phi_{18|L(\domain_{18})})$. The
	leftmost column gives all ECA neighborhood values in lexicographical order,
	and each subsequent column is the output of the neighborhoods for the
	specified dynamic or subdynamic. Symbol $\excluded$ indicates a lookup
	table element excluded from the respective subdynamic.
	}
\label{tab:rule18}
\end{table}

Having defined and described rule 18's domain $\domain_{18}$ allows us to
explain its linearization to rule 90: Keeping only elements of
$\lut(\phi_{18})$ that are also in $\lut(\phi_{90})$ gives the linearization
$\lut(\phi_{18 \leftrightarrow 90})$ of rule 18 to rule 90. This is shown in
Table~\ref{tab:rule18}. In contrast, keeping only elements of $\lut(\phi_{18})$
if the neighborhood $\neighborhood_i$ of that element belongs to the
$0$-$\Sigma$ language of $\domain_{18}$ gives the restriction $\lut(\phi_{18 |
L(\domain_{18})})$ of $\phi_{18}$ to its domain. As shown in
Table~\ref{tab:rule18}, this subdynamic excludes the neighborhoods
$\neighborhood \in \{111, 110, 011\}$. Table~\ref{tab:rule18} shows that the
only two elements that differ between $\lut(\phi_{90})$ and $\lut(\phi_{18})$
have neighborhoods $\neighborhood \in \{110, 011\}$. Thus, $\lut(\phi_{18 |
L(\domain_{18})}) \subset \lut(\phi_{18 \leftrightarrow 90})$. Moreover, as
$\domain_{18}$ is a stochastic domain with recurrence time $\widehat{p} = 1$,
we find that rule 18 restricted to its domain linearizes to rule 90 at all
powers of the lookup tables:
\begin{align*}
\lut(\phi_{18 | L(\domain_{18})}^\power) \subseteq \lut(\phi_{18 \leftrightarrow 90}^\power)
~,
\end{align*}
for $n = 1, 2, 3, \ldots$.

\subsubsection{Invariant sets of rule 90}

Historically the $0$-$\Sigma$ domain was of interest because the nonlinear
rule 18 exhibits linear behavior over $\domain_{0,\Sigma}$, since it emulates
the linear rule 90 over $\domain_{0,\Sigma}$. However, we now know that
$\domain_{0,\Sigma}$ is an invariant set of rule 18 and $\lut(\phi_{18 |
L(\domain_{0,\Sigma})}) \subset \lut(\phi_{18 \leftrightarrow 90})$. This
means $\domain_{0,\Sigma}$ is also an invariant set of rule 90. 
 
Since $\lut(\phi_{90})$ is additive, so is $\lut(\phi_{90 |
L(\domain_{0,\Sigma})})$. And, as described above, there are three
neighborhoods $\neighborhood \in \{111, 110, 011\}$ in $\lut(\phi_{90})$ that
are not in $\lut(\phi_{90 | L(\domain_{0,\Sigma})})$. So, starting from
$\lut(\phi_{90 | L(\domain_{0,\Sigma})})$, which is additive, there are three
unconstrained outputs, one for each $\neighborhood \in \{111, 110, 011\}$, to
fill in to create an ECA lookup table that has $\domain_{0,\Sigma}$ as a linear
invariant set. This is shown graphically in \cref{tab:zerowildcard}. The
neighborhoods are ordered there so that the top five neighborhoods are those in
$L(\domain_{0,\Sigma})$ and the bottom three are those that are not. 
 
\begin{table}
\begin{tabular}{c c c | c | c | c | c | c | c | c | c }
\multicolumn{3}{c|}{$\eta$} & $\phi_{90}$ & $\phi_{18}$ & $\phi_{26}$ &$\phi_{82}$ & $\phi_{146}$ & $\phi_{154}$ & $\phi_{210}$ & $\phi_{218}$ \\
\hline
1&0&1 & 0&0&0&0&0&0&0&0\\
1&0&0 & 1&1&1&1&1&1&1&1\\
0&1&0 & 0&0&0&0&0&0&0&0\\
0&0&1 & 1&1&1&1&1&1&1&1\\
0&0&0 & 0&0&0&0&0&0&0&0\\
\hline
1&1&1 & 0&0&0&0&1&1&1&1\\
1&1&0 & 1&0&0&1&0&0&1&1\\
0&1&1 & 1&0&1&0&0&1&0&1
\end{tabular}
\caption{Lookup tables for rule 90 ($\phi_{90}$) and the seven nonlinear rules,
	$\phi_\alpha$, $\alpha \in \{18, 26, 82, 146, 154, 210, 218\}$, that also
	have the $\domain_{0, \Sigma}$ domain invariant set. The first five rows
	correspond to the neighborhoods that belong to the domain language
	$L(\domain_{0,\Sigma})$. Since all eight rules have $\domain_{0,\Sigma}$ as
	a domain, the output for these five neighborhoods in
	$L(\domain_{0,\Sigma})$ are the same. The bottom three rows are the
	neighborhoods not in $L(\domain_{0,\Sigma})$. The eight rules in this table
	are all possible $2^3$ output assignments for these three remaining
	neighborhoods.
	}
\label{tab:zerowildcard}
\end{table}
 
Rule 18 is just one of seven nonlinear rules that have $\domain_{0,\Sigma}$ as
an invariant set that linearizes to rule 90: $\lut(\phi_{\alpha |
L(\domain_{0,\Sigma})}^\power) \subseteq \lut(\phi_{\alpha \leftrightarrow
90}^\power)$, for $n=1,2,3,\ldots$ and $\alpha \in \{18, 26, 82, 146, 154, 210,
218\}$. Though Rule 146 emulating rule 90 over $\domain_{0,\Sigma}$ was pointed
out in Ref. \cite{Jen90b}, to our knowledge the analysis here is the first
identification of the linear $\domain_{0,\Sigma}$ domain in the nonlinear ECAs
26, 82, 154, 210, and 218. This is likely because $\domain_{0,\Sigma}$ does not
appear to be a dominant behavior of rules 26, 82, 154, 210, and 218 from random
initial conditions. In fact, simple stationary or oscillatory behaviors seem to
be the dominant attractors for these rules.
 
\begin{figure}[htp]
 \centering
 \includegraphics{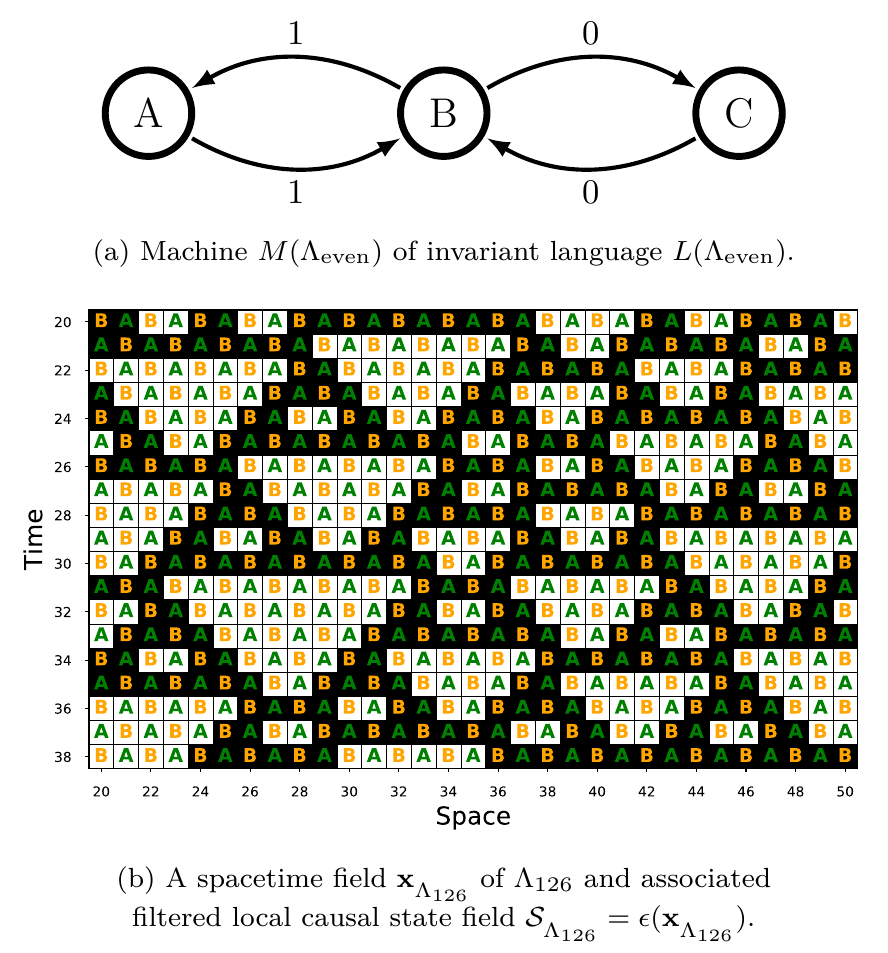}
\caption{(a) Machine $M(\domain_{\mathrm{even}})$ for the invariant language
	$L(\domain_{\mathrm{even}})$ of the rule 126 even domain. (b) Sample
	spacetime field $\stfield{\domain_{126}}{}$ (black and white squares) of
	the even domain $\domain_{\mathrm{even}}$ of rule 126 with the associated
	\localstate field
	$\causalfield{\domain_{126}}{}=\epsilon(\stfield{\domain_{126}}{})$ (green
	and orange letters) superimposed.
	}
\label{fig:domain126}
\end{figure}

Reference \cite{Jen90b} also reported on the nonlinear rule 126 emulating rule
90. This was not over $\domain_{0,\Sigma}$ though. Instead it was over a
different invariant set, that we call the \emph{even} domain
$\domain_{\mathrm{even}}$. This domain also consists of a single temporal
phase, which is the sofic shift that contains only even blocks of $1$s and
$0$s. The machine $M(\domain_{\mathrm{even}})$ for this domain is shown in
\cref{fig:domain126}. A sample domain spacetime field
$\stfield{\domain_{\mathrm{even}}}{}$, evolved from a domain configuration
initial condition, is shown in Fig.~\ref{fig:domain126} with the associated
\localstate field $\causalfield{\domain_{\mathrm{even}}}{} =
\epsilon(\stfield{\domain_{\mathrm{even}}}{})$ superimposed. Interestingly,
though $\domain_{\mathrm{even}}$ and $\domain_{0,\Sigma}$ have different
invariant spatial shifts, the resulting spacetime shifts of their orbits have
the same generalized symmetries, as captured by the local causal states.

\begin{table}
\begin{tabular}{c c c | c | c | c | c}
\multicolumn{3}{c|}{$\eta$} & $\phi_{90}(\eta)$ & $\phi_{126}(\eta)$ & $\phi_{126 \leftrightarrow 90}(\eta)$ &$\phi_{126 | L(\domain_{\mathrm{even}})}(\eta)$ \\
\hline
1 & 1 & 1 & 0 & 0 & 0 & 0\\
1 & 1 & 0 & 1 & 1 & 1 & 1\\
1 & 0 & 1 & 0 & 1 & $\excluded$ & $\excluded$ \\
1 & 0 & 0 & 1 & 1 & 1 & 1\\
0 & 1 & 1 & 1 & 1 & 1 & 1\\
0 & 1 & 0 & 0 & 1 & $\excluded$ & $\excluded$\\
0 & 0 & 1 & 1 & 1 & 1 & 1\\
0 & 0 & 0 & 0 & 0 & 0 & 0
\end{tabular}
\caption{Lookup tables for rule 90 ($\phi_{90}$) and rule 126 ($\phi_{126}$) as
	well as for rule 126 linearized to rule 90 ($\phi_{126 \leftrightarrow 90}$)
	and rule 126 restricted to its domain ($\phi_{126|L(\domain_{\mathrm{even}})})$.
	Same format as in Table \ref{tab:rule18}.
	}
\label{tab:rule126}
\end{table}

The lookup table for rule 126 is compared with that of rule 90 in
Table~\ref{tab:rule126}, as well as its linearization to rule 90 and its
restriction to $\domain_{\mathrm{even}}$. From this we can see that
$\lut(\phi_{126 | L(\domain_{\mathrm{even}})}) = \lut(\phi_{126 \leftrightarrow
90})$. As with $\domain_{0,\Sigma}$ and rule 18, this means
$\domain_{\mathrm{even}}$ is also an invariant set of rule 90. Thus, rule 126
linearizes to rule 90 over $\domain_{\mathrm{even}}$ at all powers:
\begin{align*}
\lut(\phi_{126 | L(\domain_{\mathrm{even}})}^\power) \subseteq \lut(\phi_{126 \leftrightarrow 90}^\power)
~,
\end{align*}
for $\power = 1, 2, 3, \ldots$.

Following the same combinatorics as with $\domain_{0,\Sigma}$, we see from
\cref{tab:rule126} that there are only two neighborhoods not in
$L(\domain_{\mathrm{even}})$. And so, there are $2^2$ ECA rules with
$\domain_{\mathrm{even}}$ as a domain invariant set. Two of these are rule 90
and rule 126; the other two are rule 94 and rule 122. Rule 122 is qualitatively
similar to rule 126 over random initial conditions, while rule 94 generically
settles into a fixed-point orbit.

Before moving on, clarification is in order. We said that $\domain_{0,\Sigma}$
and $\domain_{\mathrm{even}}$ are domain invariant sets of rule 90. Formally,
from \cref{defn:dpiddomain}, $\Phi_{90}$ is a factor map from $\domain$ to
$\domain$ for both of these domains. More specifically, this is true for
\emph{every} power of $\Phi_{90}$: $\Phi^\power_{90}: \domain \rightarrow
\domain$ for all $\power = 1, 2, 3, \ldots$. This is also holds for all the
nonlinear rules just discussed that also have one of these domain invariant
sets. Thus, these rules \emph{emulate} rule 90 over their domain. That is, for
all of these nonlinear $\phi_{\alpha}$, any orbit starting from an initial
configuration $\widehat{\state}$ in the appropriate domain $\domain$ is the
same if evolved under $\Phi_{\alpha}$ or $\Phi_{90}$:
\begin{align*}
\{\widehat{\state}, \Phi_{\alpha}(\widehat{\state}),
& \Phi^2_{\alpha}(\widehat{\state}), \Phi^3_{\alpha}(\widehat{\state}),
\ldots \} \\
   & \qquad =
	\{\widehat{\state}, \Phi_{90}(\widehat{\state}),
	\Phi^2_{90}(\widehat{\state}), \Phi^3_{90}(\widehat{\state}), \ldots \}
 ~.
\end{align*}

\subsubsection{Rule 22}

The next example we explore is the enigmatic rule 22 \cite{Grass86b}, which
exhibits a more general notion of linearization. Using symbolic manipulation
methods, Crutchfield and McTague used the FME analysis to discover this ECA's
domain \cite{Crut02a}.

\begin{figure}[htp]
  \centering
  \includegraphics{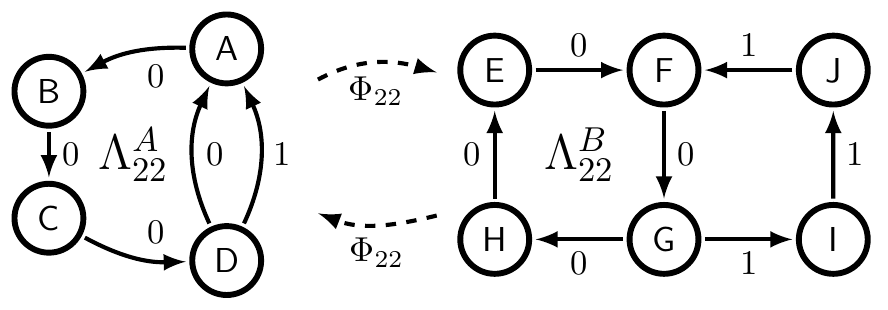}
\caption{Machine $M(\domain_{22})$ for rule 22's domain $\domain_{22}$, which
	has two distinct temporal phases: $\domain_{22}^A = \Phi_{22}
	(\domain_{22}^B)$ and $\domain_{22}^B = \Phi_{22} (\domain_{22}^A)$.
	}
\label{fig:domain22machine}
\end{figure}

Rules dominated by a stochastic symmetry domain, such as rule 22, are sometimes
referred to as ``chaotic'' CAs. As such, it is typically challenging to extract
meaningful structures purely from visually inspecting spacetime fields. So,
while not visually apparent, the domain underlying rule 22 is rather more
complex than the previous ones. Much of rule 22's mystery stems from its
complex domain.

\begin{figure}[htp]
\centering
\includegraphics[width = \columnwidth]{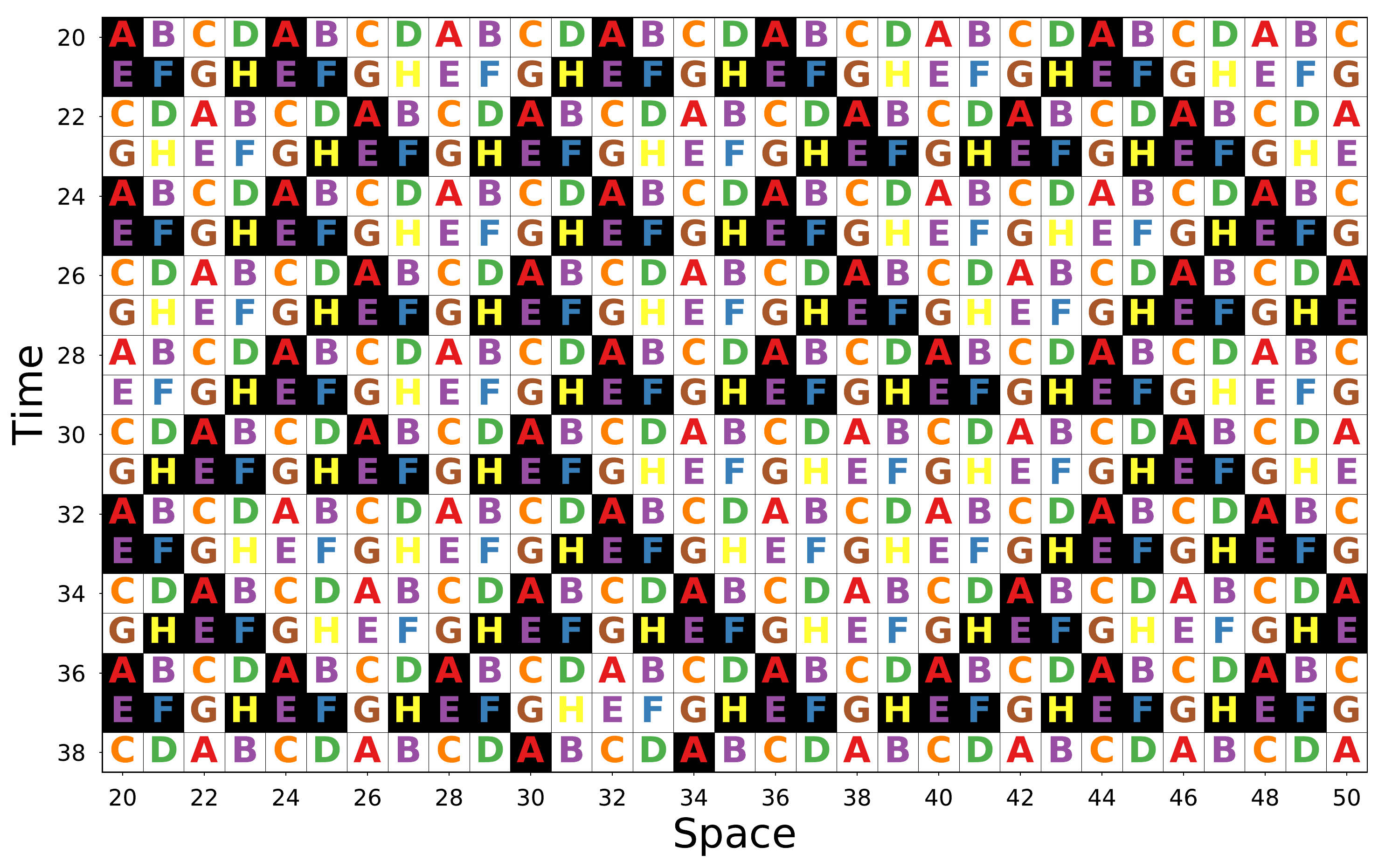}
\caption{Filtered spacetime field $\stfield{\domain_{22}}{}$ (white and
	black squares) of the rule 22 domain $\domain_{22}$ with the associated
	\localstate field
	$\causalfield{\domain_{22}}{}=\epsilon(\stfield{\domain_{22}}{})$ (colored
	letters) superimposed.}
\label{fig:domain22}
\end{figure}

The domain of rule 22 is comprised of two temporal phases, $\domain_{22} =
\{\domain_{22}^A , \domain_{22}^B\}$. The machine representation
$M(\domain_{22})$ of $\domain_{22}$ is shown in Fig.~\ref{fig:domain22machine}.
The two components correspond to the irreducible sofic shifts $\domain_{22}^A $
and  $\domain_{22}^B$. A sample spacetime field $\stfield{\domain_{22}}{}$ of
$\domain_{22}$ is shown in Fig.~\ref{fig:domain22}, with the associated
\localstate field $\causalfield{\domain_{22}}{} =
\epsilon(\stfield{\domain_{22}}{})$ superimposed on top. There are two distinct
spatial tilings of the \localstates, {\sf ABCD} and {\sf EFGH}, associated with
$\domain_{22}^A $ and $\domain_{22}^B$, respectively, giving a recurrence time
$\widehat{p} = 2$. The spacetime translation invariance of
$\causalfield{\domain_{22}}{}$ gives a spatial period of $s = 4$ and temporal
period $p = 4$.

Since $\domain_{22}$ has two temporal phases, care must be taken when
discussing its invariance and linearization. Reference \cite{Crut02a}'s FME
analysis established that $\domain_{22}^A = \Phi_{22}(\domain_{22}^B)$ and
$\domain_{22}^B = \Phi_{22}(\domain_{22}^A)$. Thus, $\Phi_{22}$ is a factor map
from each phase to itself only at the second power: $\Phi^2_{22}:
\domain_{22}^A \rightarrow \domain_{22}^A$ and $\Phi^2_{22}: \domain_{22}^B
\rightarrow \domain_{22}^B$. It is not surprising then that $\lut(\phi_{22 |
L(\domain_{22})})$ is not additive at the first power, but at the second power.
Specifically, $\lut(\phi^2_{22 | L(\domain_{22})})$ linearizes again to rule
90. In fact, exact symbolic calculation finds this linearization occurs at all
even powers:
\begin{align*}
\lut(\phi_{22| L(\domain_{22})}^{2\power}) \subseteq \lut(\phi_{22 \leftrightarrow 90}^{2\power})
~,
\end{align*}
for $\power = 1, 2, 3, \ldots$.

To be clear, $\lut(\phi_{22 | L(\domain_{22})})$ is constructed by keeping only
neighborhoods that are in the language $L(\domain_{22})$, which is the union
$L(\domain_{22}) = L(\domain_{22}^A) \cup L(\domain_{22}^B)$ of the temporal
phase languages. Since $\lut(\phi^{2\power}_{22 | L(\domain_{22})})$ is
additive, then its subsets $\lut(\phi^{2\power}_{22 | L(\domain^A_{22})})$ and
$\lut(\phi^{2\power}_{22 | L(\domain^B_{22})})$ are also additive. That is,
$\lut(\phi^{2\power}_{22 | L(\domain^A_{22})}) \subseteq \lut(\phi_{22
\leftrightarrow 90}^{2\power})$ and similarly for phase $B$. Therefore
$\domain^A_{22}$ and $\domain^B_{22}$ are invariant sets of $\Phi^2_{90}$.
However, we cannot say $\domain_{22}$ is a domain of rule 90 because
$\domain^A_{22} \neq \Phi_{90}(\domain^B_{22})$ and $\domain^B_{22} \neq
\Phi_{90}(\domain^A_{22})$. Thus, rule 22 over its domain does not fully
emulate rule 90, they only agree every other time step. Given similar
terminology used elsewhere, we may call $\domain_{22}$ a \emph{quasidomain} of
rule 90. 

Table~\ref{tab:rule22} gives the $2^{\mathrm{nd}}$ power of the rule 22 lookup
table as well as that for rule 90. It also gives the linearization to rule 90
as well as the restriction of rule 22 to $\domain_{22}$ at the
$2^{\mathrm{nd}}$ power, where the first linearization of this subdynamic
occurs.

As with $\domain_{0,\Sigma}$ and $\domain_{\mathrm{even}}$, rule 22's domain
linearizes to the additive rule 90. Rule 90 produces the mod-$2$ Pascal
triangle spacetime patterns characteristic of many chaotic CAs. It is well
known that the sum-mod-$2$ of the neighborhood outer bits is the mechanism that
generates these patterns. Since $\domain_{0,\Sigma}$ and
$\domain_{\mathrm{even}}$ are a subset of rule 90's behaviors they also exhibit
the mod-$2$ Pascal triangles, as can be seen in \cref{fig:domain18,fig:domain126}. Since the discovery of rule 22's domain, it was known
that it produces similar Pascal triangle patterns. Though these are not exactly
the same as rule 90's, since $\phi_{22|L(\domain_{22})}$ only emulates rule 90
every other time step. However, as far as we are aware, rule 22's domain
linearization to rule 90 at every even power here is the first report of such a
mechanism.

\begin{table}
\begin{tabular}{c c c c c | c | c | c | c}
\multicolumn{5}{c|}{$\eta^2$} & $\phi_{90}^2(\eta^2)$ & $\phi_{22}^2(\eta^2)$ & $\phi_{22 \leftrightarrow 90}^2(\eta^2)$ &$\phi^2_{22 | L(\domain_{22})}(\eta^2)$ \\
\hline
1 & 1 & 1 & 1 & 1    &    0  &  0    &    0  &  $\excluded$\\
1 & 1 & 1 & 1 & 0    &    1  &  0    &    $\excluded$  &  $\excluded$\\
1 & 1 & 1 & 0 & 1    &    0  &  0    &    0  &  0\\
1 & 1 & 1 & 0 & 0    &    1  &  1    &    1  &  1\\
1 & 1 & 0 & 1 & 1    &    0  &  0    &    0  &  0\\
1 & 1 & 0 & 1 & 0    &    1  &  1    &    1  &  $\excluded$\\
1 & 1 & 0 & 0 & 1    &    0  &  0    &    0  &  $\excluded$\\
1 & 1 & 0 & 0 & 0    &    1  &  1    &    1  &  1\\
1 & 0 & 1 & 1 & 1    &    0  &  0    &    0  &  0\\
1 & 0 & 1 & 1 & 0    &    1  &  0    &    $\excluded$  &  $\excluded$\\
1 & 0 & 1 & 0 & 1    &    0  &  1    &    $\excluded$  &  $\excluded$\\
1 & 0 & 1 & 0 & 0    &    1  &  0    &    $\excluded$  &  $\excluded$\\
1 & 0 & 0 & 1 & 1    &    0  &  0    &    0  &  $\excluded$\\
1 & 0 & 0 & 1 & 0    &    1  &  0    &    $\excluded$  &  $\excluded$\\
1 & 0 & 0 & 0 & 1    &    0  &  0    &    0  &  0\\
1 & 0 & 0 & 0 & 0    &    1  &  1    &    1  &  1\\
0 & 1 & 1 & 1 & 1    &    1  &  0    &    $\excluded$  &  $\excluded$\\
0 & 1 & 1 & 1 & 0    &    0  &  0    &    0  &  0\\
0 & 1 & 1 & 0 & 1    &    1  &  0    &    $\excluded$  &  $\excluded$\\
0 & 1 & 1 & 0 & 0    &    0  &  1    &    $\excluded$  &  $\excluded$\\
0 & 1 & 0 & 1 & 1    &    1  &  1    &    1  &  $\excluded$\\
0 & 1 & 0 & 1 & 0    &    0  &  0    &    0  &  $\excluded$\\
0 & 1 & 0 & 0 & 1    &    1  &  0    &    $\excluded$  &  $\excluded$\\
0 & 1 & 0 & 0 & 0    &    0  &  0    &    0  &  0\\
0 & 0 & 1 & 1 & 1    &    1  &  1    &    1  &  1\\
0 & 0 & 1 & 1 & 0    &    0  &  1    &    $\excluded$  &  $\excluded$\\
0 & 0 & 1 & 0 & 1    &    1  &  0    &    $\excluded$  &  $\excluded$\\
0 & 0 & 1 & 0 & 0    &    0  &  0    &    0  &  0\\
0 & 0 & 0 & 1 & 1    &    1  &  1    &    1  &  1\\
0 & 0 & 0 & 1 & 0    &    0  &  0    &    0  &  0\\
0 & 0 & 0 & 0 & 1    &    1  &  1    &    1  &  1\\
0 & 0 & 0 & 0 & 0    &    0  &  0    &    0  &  0
\end{tabular}
\caption{Second-order lookup tables for rule 90 ($\phi^2_{90}$) and rule 22
	($\phi^2_{22}$), as well as for $\phi^2_{22}$ linearized to $\phi^2_{90}$
	($\phi^2_{22 \leftrightarrow 90}$) and $\phi^2_{22}$ restricted to its
	domain ($\phi^2_{22|L(\domain_{22})})$. The leftmost column gives all
	second-order ECA neighborhood values (that is, all radius-2 neighborhood
	values) in lexicographical order. Each subsequent column is the output of
	the neighborhoods for the specified dynamic or subdynamic. The symbol
	$\excluded$ indicates that lookup table element is excluded from the
	respective subdynamic.
	}
\label{tab:rule22}
\end{table}

\begin{figure}[htp]
 \includegraphics{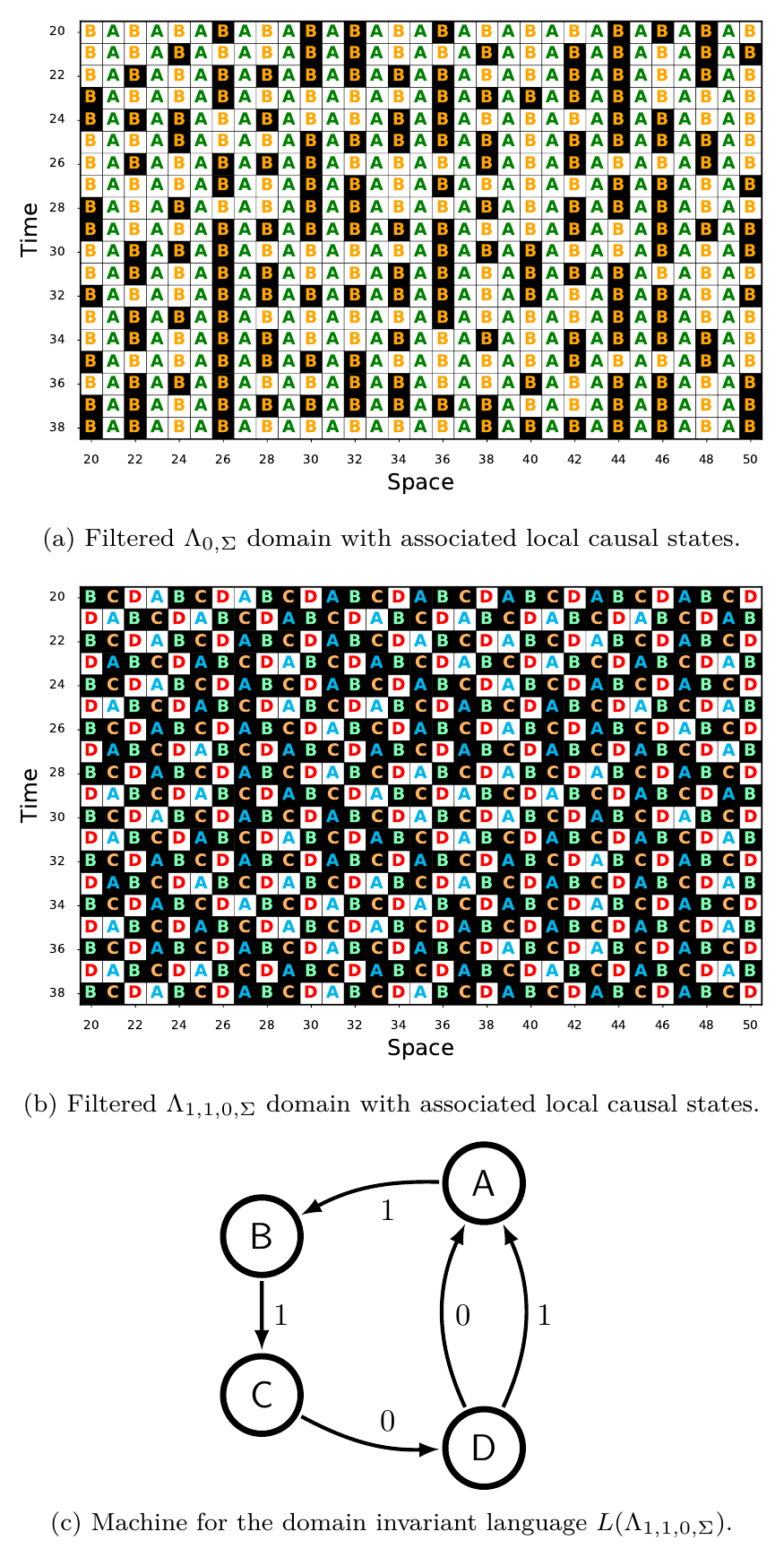}
\caption{Filtered spacetime fields of the two domains of the
	$(\MeasAlphabet = \{0,1\}, \radius = 2)$ CA rule 2614700074. Format and
	notation similar to previous such diagrams. (a) The $\domain_{0,\Sigma}$
	domain has the same invariant set language of ECA rule 90. Its machine is
	shown in Fig.~\ref{fig:domain18}(a)). (b) The $\domain_{1,1,0,\Sigma}$
	domain. (c) Machine for $L(\domain_{1,1,0,\Sigma})$.
	}
\label{fig:twodomains}
\end{figure}

\subsubsection{A nonelementary CA}

\Cref{sec:exploring} A showed that additive dynamics always implies a domain
invariant set and conjectured that additive subdynamics over a language
$L(\ShiftSpace)$ potentially also implies $\ShiftSpace$ is an invariant set.
\Cref{sec:exploring} B showed that explicit symmetry domains always imply an
additive subdynamic, connecting the periodic orbits of these domains with the
additive identity rule. So far in \cref{sec:exploring} C we have seen many
examples of ECA hidden symmetry domains that similarly are associated with an
additive subdynamic, in this case connected to the additive rule 90. However,
now we come to an example in the class $(\MeasAlphabet = \{0,1\}, \radius = 2)$
that breaks the connection between domains and additive subdynamics. It
possesses a domain that admits no linearizations.

The CA in question is radius-2 rule 2614700074, named according to the same
numbering scheme used for ECAs. This was previously studied by Crutchfield and
Hanson \cite{Crut93a}. They designed it to have the $\domain_{0, \Sigma}$
domain along with another structurally distinct domain---the
$\domain_{1,1,0,\Sigma}$ domain. This domain has a single temporal phase
consisting of the sofic shift $\ShiftSpace_{1,1,0,\Sigma}$ with strings of the
form $\cdots110\Sigma110\Sigma110\Sigma\cdots$, where $\Sigma$ is a wildcard
that can be either $1$ or $0$. The machine for $\domain_{1,1,0,\Sigma}$ is
shown in Fig.~\ref{fig:twodomains}(c).

Reference~\cite{Crut93a} showed that $\domain_{0,\Sigma}$ and
$\domain_{1,1,0,\Sigma}$ have distinct statistical signatures. Here, we
investigate these differences via \localstates. Filtered spacetime fields for
the $\domain_{0,\Sigma}$ and $\domain_{1,1,0,\Sigma}$ domains are shown in
Fig.~\ref{fig:twodomains}(a) and (b), respectively. In each, as above, the
colored letters represent the \localstate label at each site. These \localstate
overlay diagrams clearly demonstrate that the domains have different spacetime
symmetry groups. $\domain_{1,1,0,\Sigma}$ has recurrence time $\widehat{p} =
1$, temporal period $p = 2$, and spatial period $4$, while $\domain_{0,\Sigma}$
has recurrence time $\widehat{p}=1$, temporal period $p=1$, and spatial period
$s=2$ for rule 2614700074.

Recall that $\domain_{0,\Sigma}$ is a domain invariant set of ECA rule 90. It
is thus also a domain invariant set of $\Phi^2_{90}$, which is itself a CA in
the class $(\MeasAlphabet=\{0,1\}, R=2)$: $\mathbf{a} =(1,0,0,0,1)$. From this
we can understand the $\domain_{0,\Sigma}$ domain of rule 2614700074 from the
same combinatorial perspective as $\domain_{0,\Sigma}$ with rule 18. The
restriction $\lut(\phi^2_{90 | L(\domain_{0,\Sigma})})$ leaves several output
values unconstrained for assignment to construct a full CA lookup table. One
such assignment gives rule 2614700074. As such, we find that rule 2614700074
linearizes to $\mathbf{a} =(1,0,0,0,1)$ at all powers:
\begin{align*}
\lut(\phi^\power_{2614700074 | L(\domain_{0,\Sigma})}) \subset \lut(\phi^\power_{2614700074 \leftrightarrow \mathbf{a} =(1,0,0,0,1)})
,
\end{align*}
for $\power = 1, 2, 3, \ldots$.

This connection to rule 90 also explains why $\domain_{0,\Sigma}$ has temporal
period $p=2$ for rule 18, but temporal period $p=1$ for CA 2614700074. The
local causal state field of $\domain_{0,\Sigma}$ for rule 18 has a checkerboard
symmetry. If one starts in state $A$ at some spacetime point and moves forward
one time step (i.e., applying $\Phi_{90}$) one arrives at state $B$. However,
starting in state $A$ and moving forward two time steps (i.e., applying
$\Phi^2_{90}$) one ends in state $A$ again.

The $\domain_{1,1,0,\Sigma}$ domain of rule 2614700074, in contrast to all
other examples we have seen, does not appear to have any linearizations. From
all examples we have seen and know of so far, we would expect rule 2614700074
to linearize over $\domain_{1,1,0,\Sigma}$ at its first power since it is a
hidden symmetry domain with recurrence time $\widehat{p}=1$:
$\Phi_{2614700074}: \ShiftSpace_{1,1,0,\Sigma} \rightarrow
\ShiftSpace_{1,1,0,\Sigma}$. \Cref{app:nonlineardomain} demonstrates that
$\lut(\phi_{2614700074 | L(\domain_{1,1,0,\Sigma})})$ cannot be additive. We
also algorithmically checked for all possible linearizations of
$\lut(\phi^\power_{2614700074 | L(\domain_{1,1,0,\Sigma})})$ for $\power =
1,2,3,4$, finding none. For each $\power \in \{1,2,3,4\}$ we constructed the
linearization $\lut(\phi^\power_{2614700074 \leftrightarrow \mathbf{a}})$ for
all possible length $2\power \radius + 1$ coefficient vectors $\mathbf{a}$ and
found $\lut(\phi^\power_{2614700074 | L(\domain_{1,1,0,\Sigma})})$ to be a
subset of none of the linearizations.

\section{Conclusion}

The most basic ingredient of a cellular automaton, its lookup table, could not
be simpler---a finite number of possible inputs are enumerated with their
outputs explicitly specified. However, the overlapping interactions that occur
when applying this simple lookup table synchronously for simultaneous global
update of spatial configurations conspires to produce arbitrarily complex
behaviors. The emergent complexity enshrouds a cellular automaton's simplicity,
making it difficult to answer seemingly basic questions. Specifying a lookup
table $\phi$ determines the global update $\Phi$. Given a lookup table for
$\phi$, and hence $\Phi$, what invariant sets are induced by $\Phi$ in the
state space $\FullShift$? In contrast with low-dimensional dynamical systems,
the states $\state \in \FullShift$ of spatially-extended dynamical systems like
CAs possess internal structure and live in infinite dimensions. For a given
invariant set $\domain \subseteq \FullShift$, is there a unifying structure in
the states $\state \in \domain$?  Moreover, is there spacetime structure in the
orbits of sequential states in $\domain$? 

Our investigations provide several inroads to these questions, but they also
highlight the challenge presented by complex spatially-extended dynamical
systems and their emergent behaviors. While domain invariant sets appear to
strongly correspond to the spacetime symmetries revealed by the \localstates in
the orbit flows along the invariant sets, relating the invariant spatial shift
spaces of \cref{defn:dpiddomain} with the resulting spacetime shift spaces of
\cref{defn:lcsdomain} and their generalized symmetries remains unsolved.  Since
$\Phi$ is deterministic, the spacetime shift space that results from a given
spatial shift space $\ShiftSpace$ is uniquely determined by $\Phi$. This is not
to say, though, as is often assumed, that the spacetime shift space trivially
follows from $\Phi$ applied to $\ShiftSpace$. In fact, we still do not know how
to fully characterize the spacetime shift space of orbits that follow from
hidden symmetry domain invariant spatial shift spaces. In particular, we do not
know how to properly define the domain temporal period for these spaces from
their invariant spatial shift spaces. Such difficulties in understanding the
spacetime shift spaces of hidden symmetry domains is perhaps one of the
clearest examples of the fallacy of the ``constructionist'' hypothesis that
often accompanies reductionism \cite{Ande72a}. Tackling this will be a focus of
subsequent investigations.

Beyond characterizing the spacetime shift spaces of domains, there remains the
challenge of connecting domains, both the invariant spatial shift spaces and
their resulting spacetime shift space of orbits, to the equation of motion
$\Phi$ and the lookup table $\phi$ that generates it. Why does the particular
assignment of lookup table outputs that form $\lut(\phi_{18})$ generate the
invariant set $\domain_{0,\Sigma}$? Why should $\lut(\phi_{18 |
L(\domain_{0,\Sigma})})$ be additive when $\lut(\phi_{18})$ is not? We have
shown that this linear behavior of the nonlinear rule 18 actually follows from
the combinatorics of $\domain_{0,\Sigma}$ being an invariant set of the
additive rule 90. In fact, this is the mechanism behind every known stochastic
ECA domain, including the enigmatic rule 22. Is this due to historical focus on
rule 90? Or, is rule 90 particularly special? Beyond ECAs, we know from the
$\domain_{1,1,0,\Sigma}$ domain of the $R=2$ rule 2614700074 that this is not
the only mechanism for generating hidden symmetry domains. Hidden symmetry
domains need not be associated with an additive subdynamic. One possible path
forward could be through the \emph{partial permutivity} outlined in Ref.
\cite{elor93a}; the permutive subalphabets of examples $1.2$ and $1.3$ there
correspond to the domains of ECAs 18 and 22, respectively. 

Perhaps in sixteenth and seventeenth centuries---the burgeoning days of
celestial mechanics---one could take for granted that knowing the equations of
motion was tantamount to knowing the system and its behavior. Those days have
long passed. Today, we appreciate that having the Navier-Stokes equations of
hydrodynamics in hand does not translate into understanding emergent coherent
structures, such as fluid vortices. The preceding recounted this lesson yet
again. Even systems as ``simple'' as elementary cellular automata continue to
hold surprises.

\section*{Acknowledgments}
\label{sec:acknowledgments}

The authors thank Greg Wimsatt for helpful discussions and feedback. As a
faculty member, JPC thanks the Santa Fe Institute and the Telluride Science
Research Center for their hospitality during visits. This material is based
upon work supported by, or in part by, the U.S. Army Research Laboratory and
the U. S. Army Research Office under contract W911NF-13-1-0390 and grant
W911NF-18-1-0028 and via Intel Corporation support of CSC as an Intel Parallel
Computing Center.

\appendix

\section{Proof of \cref{thrm:surjectivity}}
\label{app:surjectivity}

\begin{repthrm}{thrm:surjectivity}
Every nonzero linear CA $\Phi_\beta$ is a factor map from the full-$\MeasAlphabet$ shift to itself: $\Phi_\beta : \FullShift \rightarrow \FullShift$.
\end{repthrm}

\begin{prf}
For any $x \in \FullShift$ we want to show that there exists a $y \in \FullShift$ such that $x = \Phi_\beta(y)$. In other words, for a linear $\Phi_\beta$ there is always a right inverse $\widetilde{\Phi}_\beta$ such that $\Phi_\beta\bigl(\widetilde{\Phi}_\beta\left(x\right)\bigr) = x$. 

First, decompose the string $x$ into several `basis' strings as follows; for each index $i$ in $x$ such that $x_i = 1$, create a basis string $x^i$ for which $x^i_i = 1$ and $x^i_j=0$ for all $j \neq i$. With this we have $x = \sum_i x^i$, which is shorthand for performing $x_j = \sum_i x^i_j \; (\mathrm{mod} \; 2)$ at each index $j$ of $x$. 

If we can show the basis strings $x^i$ always have a pre-image $y^i$, $x^i =
\Phi_\beta(y^i)$, then it follows from linear superposition that $y = \sum_i
y^i$ is the pre-image of $x$. If $x = \sum_i x^i = \sum_i \Phi_\beta(y^i)$,
then using superposition, \cref{eqn:linearity}, we have $\sum_i \Phi_\beta(y^i)
= \Phi_\beta ( \sum_i y^i)$. Thus, for any $x$ there is a $y$ such that $\Phi_\beta(y) = \Phi(\sum_i y^i) = \sum_i \Phi_\beta (y^i) = \sum_i x^i = x$.

We now need to show that for any basis string $x^i$ with a single $1$ there is
always a pre-image $y^i$. This follows from additivity of $\phi_\beta$. Since
we are considering nonzero linear CAs there is at least one coefficient $a_i =
1$ in the additivity coefficient vector $\mathbf{a}$. In fact, it simplifies
things to only consider CAs that have one or both of the outer bits of the
neighborhood with a nonzero coefficient: $a_{-R} = 1$ or $a_{R} = 1$ or both.
Any rule where this is not the case is equivalent to one that is; e.g., the
radius $R=2$ CA $\mathbf{a} = (0,1,0,1,0)$ is equivalent to the radius $R=1$ CA
$\mathbf{a} = (1,0,1)$. (The only exception is the identity rule with $a_0 =1$
and $a_i = 0$ for all $i \in \{-R, \ldots, 0, \ldots, R\} \backslash \{0\}$,
but clearly every image is its own pre-image for the identity rule, so it is
surjective). 

Chose $i$ to be either $-R$ or $R$ such that $a_i =1$. The neighborhood that
has $x_i = 1$ and $x_j = 0$ for $j \in \{-R, \ldots, 0, \ldots, R\}$ and $j
\neq i$ will necessarily output $1$. To find the string that outputs the
3-block $010$ there are three neighborhoods to consider, $\neighborhood_{-1}$,
$\neighborhood_0$, and $\neighborhood_{1}$. For the central neighborhood
$\neighborhood_0$, chose the same as above to still output $1$. Neighborhood
$\neighborhood_{-1}$ is to the left of $\neighborhood_0$ and $\neighborhood_1$
to the right. The outer neighborhoods overlap with $\neighborhood_0$ and thus
share all but one entry: $\neighborhood_{-1}[-(R-1), \ldots, R] =
\neighborhood_0[-R, \ldots, R-1]$ and $\neighborhood_1[-R, \ldots, R-1] =
\neighborhood_0[-(R-1), \ldots, R]$. Thus, we need only fill in
$\neighborhood_{-1}[-R]$ and $\neighborhood_1[R]$. If $i=-R$, then
$\neighborhood_0[-R] = 1$ and $\neighborhood_{-1}[-(R-1)] = 1$. And so, for
$\neighborhood_{-1}$ to output a $0$ we set $\neighborhood_{-1}[-R] = 1$ if
$a_{-R+1} = 1$ or $\neighborhood_{-1}[-R] = 0$ if $a_{-R+1} = 0$. Meanwhile
$\neighborhood_1[R-1] = 0$, so for $\neighborhood_1$ to output a $0$ we must
set $\neighborhood_1[R] = 0$. If $i = R$ perform the symmetric operation: set
$\neighborhood_{-1} [-R] = 0$ and $\neighborhood_{1}[R] = 1$. 

To output the 5-block $00100$ we similarly have two new neighborhoods to
consider, $\neighborhood_{-2}$ and $\neighborhood_2$. As above, we only need to
fill in $\neighborhood_{-2}[-R]$ and $\neighborhood_{2}[R]$. If $i=-R$ we again
just set $\neighborhood_2[R] = 0$. Now, simply set $\neighborhood_{-2}[-R]$ to
either $0$ or $1$ so that $\mathbf{a} \cdot \neighborhood_{-2} = 0$. Perform
the similar symmetric construction if $i=R$. We can continue in this way to
extend out to output arbitrary blocks $\cdots000\cdots1\cdots000\cdots$ with a
single $1$. Thus, we showed how to construct a pre-image $y^i$ for any basis string $x^i = \cdots000\cdots1\cdots000\cdots$. 

All that remains is the case of the all-$0$ string as an image. Clearly,
though, from additivity, \cref{eqn:additivity}, the all-$0$ string is its own
pre-image for all linear CAs. 

If $\Phi_\beta$ is surjective over all finite blocks, as we just showed, then
$\Phi_\beta$ is necessarily surjective over $\FullShift$. Thus, $y = \sum_i y^i
= \widetilde{\Phi}_\beta(x)$ for all $x \in \FullShift$.

\hfill $\blacksquare$
\end{prf}

\section{Domain $\domain_{1,1,0,\Sigma}$ of rule 2614700074 is not linear}
\label{app:nonlineardomain}

Reference \cite{Crut93a} showed, using the FME, that $\domain_{1,1,0,\Sigma}$
is a domain invariant set of rule 2614700074. First, we check that all 5-block
words, i.e., all $R=2$ neighborhoods $\neighborhood$, are in the language
$L(\domain_{1,1,0,\Sigma})$. To do so, consider the string
$\cdots110\Sigma110\Sigma110\Sigma\cdots$ and a sliding $5$-block window over
this string. This yields the $5$-blocks $110\Sigma1$, $10\Sigma11$,
$0\Sigma110$, and $\Sigma110\Sigma$. Replacing each $\Sigma$ with realizations
$0$ and $1$ gives the neighborhoods in $L(\domain_{1,1,0,\Sigma})$, from which
we can create $\lut(\phi_{2614700074 | L(\domain_{1,1,0,\Sigma})})$, shown in
\cref{tab:twodomain}.

Now we show that there is no additivity assignment $\mathbf{a} = (a_{-2},
a_{-1}, a_0, a_1, a_2)$ such that $\phi_{2614700074 |
L(\domain_{1,1,0,\Sigma})}(\eta)$ is given by $\mathbf{a}
\stackrel{(\mathrm{mod} \; 2)}{\cdot} \neighborhood$. From the first two rows
of \cref{tab:twodomain} we see that $\phi_{2614700074 |
L(\domain_{1,1,0,\Sigma})}(11101) = 0$ and $\phi_{2614700074 |
L(\domain_{1,1,0,\Sigma})}(11100) = 1$. If $\phi_{2614700074 |
L(\domain_{1,1,0,\Sigma})}(\eta)$ is additive, this shows the right-most entry
of $\neighborhood$ must contribute to the additivity sum, and so we must have
$a_2 = 1$. Similarly, from $\phi_{2614700074 |
L(\domain_{1,1,0,\Sigma})}(01100)=0$ and $\phi_{2614700074 |
L(\domain_{1,1,0,\Sigma})}(11100)=0$ we would need $a_{-2}=1$. The
neighborhoods $11001$ and $11011$ have the same output, giving $a_1=0$.
Similarly, $10011$ and $10111$ have the same output, giving $a_0=0$, and $00110$
and $01110$ have the same output giving $a_{-1}=0$. 

Therefore, if there is an additivity assignment $\mathbf{a}$, it must be
$\mathbf{a} = (1,0,0,0,1)$. However, we can see that $\phi_{2614700074 |
L(\domain_{1,1,0,\Sigma})}(11011) = 1$, for example, and $(1,0,0,0,1)
\stackrel{(\mathrm{mod} \; 2)}{\cdot} (1,1,0,1,1) \neq 1$. And so,
$\phi_{2614700074 | L(\domain_{1,1,0,\Sigma})}(\eta)$ cannot be additive. 

\hfill $\blacksquare$
\begin{table}
\begin{tabular}{c c c c c | c}
\multicolumn{5}{c|}{$\eta$} & $\phi_{2614700074 | L(\domain_{1,1,0,\Sigma})}(\eta)$ \\
\hline
1 & 1 & 1 & 0 & 1 &   0 \\
1 & 1 & 1 & 0 & 0 &   1 \\
1 & 1 & 0 & 1 & 1 &   1 \\
1 & 1 & 0 & 0 & 1 &   1 \\
1 & 0 & 1 & 1 & 1 &   1 \\
1 & 0 & 0 & 1 & 1 &   1 \\
0 & 1 & 1 & 1 & 0 &   0 \\
0 & 1 & 1 & 0 & 1 &   1 \\
0 & 1 & 1 & 0 & 0 &   0 \\
0 & 0 & 1 & 1 & 0 &   0 \\
\end{tabular}
\caption{First-order lookup table of $(\MeasAlphabet = \{0,1\}, \radius = 2)$
	CA 2614700074 restricted to its domain $\domain_{1,1,0,\Sigma}$. For
	simplicity, all elements of $\lut(\phi_{2614700074})$ not in
	$\lut(\phi_{2614700074 | L(\domain_{1,1,0,\Sigma})})$ are not shown.
	}
\label{tab:twodomain}
\end{table}



\begin{thebibliography}{10}

\bibitem{Farm84a}
J.~D. Farmer, T.~Toffoli, and S.~Wolfram, editors.
\newblock {\em Cellular Automata, Proceedings of an Interdisciplinary
  Workshop}, Amsterdam, 1984. North-Holland Publishing Company.

\bibitem{Pack85b}
N.~H. Packard and S.~Wolfram.
\newblock Two-dimensional cellular automata.
\newblock {\em J. Stat. Physics}, 38(5--6):901--946, 1985.

\bibitem{Toff87a}
T.~Toffoli and N.~Margolis.
\newblock {\em Cellular Automata Machines: A New Environment for Modeling}.
\newblock MIT Press, Cambridge, Massachusetts, 1987.

\bibitem{Guto91b}
H.~Gutowitz.
\newblock {\em Cellular Automata: Theory and Experiment}.
\newblock Special Issues of Physica D. Bradford Books, Cambridge,
  Massachusetts, 1991.

\bibitem{Mart84a}
O.~Martin, A.~Odlyzko, and S.~Wolfram.
\newblock Algebraic properties of cellular automata.
\newblock {\em Commun. Math. Phys.}, 93:219, 1984.

\bibitem{Lind84a}
D.~A. Lind.
\newblock Applications of ergodic theory and sofic systems to cellular
  automata.
\newblock {\em Physica}, 10D:36, 1984.

\bibitem{Gras83a}
P.~Grassberger.
\newblock New mechanism for deterministic diffusion.
\newblock {\em Phys. Rev. A}, 28:3666, 1983.

\bibitem{Wolf83}
S.~Wolfram.
\newblock Statistical mechanics of cellular automata.
\newblock {\em Rev. Mod. Phys.}, 55:601, 1983.

\bibitem{Jen90b}
E.~Jen.
\newblock Exact solvability and quasiperiodicity of one-dimensional cellular
  automata.
\newblock {\em Nonlinearity}, 4:251, 1990.

\bibitem{Moor97b}
C.~Moore.
\newblock Quasilinear cellular automata.
\newblock {\em Physica D: Nonlinear Phenomena}, 103(1-4):100--132, 1997.

\bibitem{Grav11a}
J.~Gravner and D.~Griffeath.
\newblock The one-dimensional {E}xactly 1 cellular automaton: Replication,
  periodicity, and chaos from finite seeds.
\newblock {\em J. Stat. Physics}, 142(1):168--200, 2011.

\bibitem{Hans90a}
J.~E. Hanson and J.~P. Crutchfield.
\newblock The attractor-basin portrait of a cellular automaton.
\newblock {\em J. Stat. Phys.}, 66:1415 -- 1462, 1992.

\bibitem{Crut91d}
J.~P. Crutchfield.
\newblock Discovering coherent structures in nonlinear spatial systems.
\newblock In A.~Brandt, S.~Ramberg, and M.~Shlesinger, editors, {\em Nonlinear
  Ocean Waves}, pages 190--216, Singapore, 1992. World Scientific.
\newblock also appears in Complexity in Physics and Technology, R.
  Vilela-Mendes, editor, World Scientific, Singapore (1992).

\bibitem{Crut92a}
J.~P. Crutchfield and J.~E. Hanson.
\newblock Attractor vicinity decay for a cellular automaton.
\newblock {\em CHAOS}, 3(2):215--224, 1993.

\bibitem{Crut93a}
J.~P. Crutchfield and J.~E. Hanson.
\newblock Turbulent pattern bases for cellular automata.
\newblock {\em Physica D}, 69:279 -- 301, 1993.

\bibitem{Hans95a}
J.~E. Hanson and J.~P. Crutchfield.
\newblock Computational mechanics of cellular automata: An example.
\newblock {\em Physica D}, 103:169--189, 1997.

\bibitem{McTa04a}
C.~S. McTague and J.~P. Crutchfield.
\newblock Automated pattern discovery---{An} algorithm for constructing
  optimally synchronizing multi-regular language filters.
\newblock {\em Theo. Comp. Sci.}, 359(1-3):306--328, 2006.

\bibitem{Nord88a}
M.~G. Nordahl.
\newblock Limit sets of class two cellular automata.
\newblock 1988.
\newblock unpublished.

\bibitem{Rupe17b}
A.~Rupe and J.~P. Crutchfield.
\newblock Local causal states and discrete coherent structures.
\newblock {\em Chaos}, 28(7):1--22, 2018.

\bibitem{LeBr91a}
L.~Le Bruyn and M.~Van den Bergh.
\newblock Algebraic properties of linear cellular automata.
\newblock {\em Linear Algebra and its Applications}, 157:217--234, 1991.

\bibitem{Lind77a}
D.~A. Lind.
\newblock The structure of skew products with ergodic group automorphisms.
\newblock {\em Israel J. Math.}, 28(3):205--248, 1977.

\bibitem{Hopc06a}
J.~E. Hopcroft, R.~Motwani, and J.~D. Ullman.
\newblock {\em Introduction to Automata Theory, Languages, and Computation}.
\newblock Prentice-Hall, New York, third edition, 2006.

\bibitem{Lind95a}
D.~Lind and B.~Marcus.
\newblock {\em An Introduction to Symbolic Dynamics and Coding}.
\newblock Cambridge University Press, New York, 1995.

\bibitem{Broo89a}
J.~G. Brookshear.
\newblock {\em Theory of computation: {Formal} languages, automata, and
  complexity}.
\newblock Benjamin/Cummings, Redwood City, California, 1989.

\bibitem{Wolf84a}
S.~Wolfram.
\newblock Computation theory of cellular automata.
\newblock {\em Comm. Math. Phys.}, 96:15, 1984.

\bibitem{Crut12a}
J.~P. Crutchfield.
\newblock Between order and chaos.
\newblock {\em Nature Physics}, 8(January):17--24, 2012.

\bibitem{Crut88a}
J.~P. Crutchfield and K.~Young.
\newblock Inferring statistical complexity.
\newblock {\em Phys. Rev. Let.}, 63:105--108, 1989.

\bibitem{Shal98a}
C.~R. Shalizi and J.~P. Crutchfield.
\newblock Computational mechanics: Pattern and prediction, structure and
  simplicity.
\newblock {\em J. Stat. Phys.}, 104:817--879, 2001.

\bibitem{Shal03a}
C.R. Shalizi.
\newblock Optimal nonlinear prediction of random fields on networks.
\newblock {\em Discrete Mathematics \& Theoretical Computer Science}, 2003.

\bibitem{Crut91b}
J.~P. Crutchfield.
\newblock Semantics and thermodynamics.
\newblock In M.~Casdagli and S.~Eubank, editors, {\em Nonlinear Modeling and
  Forecasting}, volume XII of {\em Santa Fe Institute Studies in the Sciences
  of Complexity}, pages 317 -- 359, Reading, Massachusetts, 1992.
  Addison-Wesley.

\bibitem{John10a}
B.~D. Johnson, J.~P. Crutchfield, C.~J. Ellison, and C.~S. McTague.
\newblock Enumerating finitary processes.
\newblock arxiv.org:1011.0036.

\bibitem{Hedl69a}
G.~A. Hedlund.
\newblock Endomorphisms and automorphisms of the shift dynamical system.
\newblock {\em Theory of Computing Systems}, 3(4):320--375, 1969.

\bibitem{Note1}
Crutchfield and McTague implemented an efficient, but exhaustive search
  algorithm to solve the invariant equation using the enumerated library of
  machines of Ref. \cite {John10a}. Reference \cite {Crut02a} analyzed ECA 22
  using the approach.

\bibitem{Holc82a}
W.~M.~L. Holcombe.
\newblock {\em Algebraic Automata Theory}.
\newblock Cambridge University Press, Cambridge, 1982.

\bibitem{Grass86b}
P.~Grassberger.
\newblock Long-range effects in an elementary cellular automaton.
\newblock {\em J. Stat. Physics}, 45(1-2):27--39, 1986.

\bibitem{Crut02a}
J.~P. Crutchfield and C.~S. McTague.
\newblock Unveiling an enigma: Patterns in elementary cellular automaton 22 and
  how to discover them.
\newblock 2002.
\newblock Santa Fe Institute Technical Report.

\bibitem{Ande72a}
P.~W. Anderson.
\newblock More is different.
\newblock {\em Science}, 177(4047):393--396, 1972.

\bibitem{elor93a}
K.~Eloranta.
\newblock Partially permutive cellular automata.
\newblock {\em Nonlinearity}, 6(6):1009, 1993.

\end{thebibliography}
\end{document}